\documentclass[11pt,preprint]{aastex}

\usepackage{hyperref}

\newcommand{\xmm}{{\em XMM-Newton}}
\newcommand{\hd}{{HD~189733}}

\newcommand{\myemail}{pilli@astropa.inaf.it,ipillitt@cfa.harvard.edu}
\newcommand{\atlasurl}{www.astropa.inaf.it/~pilli/hd189733\_hst\_cos.pdf} 

\slugcomment{}

\shorttitle{FUV line variability of HD~189733 }
\shortauthors{Pillitteri et al.}

\usepackage{rotating}
\begin{document}


\title{FUV variability of HD~189733. Is the star accreting material from
its hot Jupiter?}


\author{I. Pillitteri\altaffilmark{1,2}}
\author{A. Maggio\altaffilmark{1}} 
\author{G. Micela\altaffilmark{1}} 
\author{S. Sciortino\altaffilmark{1}} 
\affil{INAF-Osservatorio Astronomico di Palermo, Piazza del Parlamento 1, 
90134, Palermo, Italy}
\author{S. J. Wolk\altaffilmark{2}} 
\affil{Harvard-Smithsonian Center for Astrophysics, Cambridge MA 02138, USA}
\author{T. Matsakos\altaffilmark{3}} 
\affil{Department of Astronomy \& Astrophysics, The University of Chicago, Chicago, IL 60637, USA}
\email{\myemail}


%

\begin{abstract}
Hot Jupiters are subject to strong irradiation from the
host stars and, as a consequence, they do evaporate. They can also
interact with the parent stars by means of tides and magnetic fields.
Both phenomena have strong implications for the evolution of these systems.
Here we present time resolved spectroscopy of HD~189733 observed with 
the Cosmic Origin Spectrograph (COS) on board to HST. The star has been observed
during five consecutive HST orbits, starting at a secondary transit of the planet
($\phi\sim0.50-0.63$). 
Two main episodes of variability of ion lines of Si, C, N and O are detected, 
with an increase of line fluxes. Si~IV lines show the highest degree of variability.
The FUV variability is a signature of enhanced activity in phase with the planet motion,
occurring after the planet egress, as already observed three times in X-rays. 
With the support of MHD simulations, we propose the following interpretation: 
a stream of gas evaporating from the planet is actively and almost steadily accreting
onto the stellar surface, impacting at $70-90\deg$ ahead of the sub-planetary point. 
\end{abstract}


\keywords{}

\section{Introduction}
The significant fraction ($\sim10\%$) of known exoplanets 
with masses of the order of the Jupiter mass and orbiting 
within few stellar radii (hot Jupiters) raised the question of
their origin and evolution.  
Because of their proximity to the parent star, these planets are strongly irradiated, 
their upper atmospheres are bloated and do evaporate 
\citep{VidalMadjar2003,Lecavelier2004,Lecavelier2010, Linsky2010,Ben-Jaffel2013}. 
It is plausible that the evaporating material forms a cometary tail and some sort
of bow shock in front of the planet due to the interaction with the stellar wind
\citep{Cohen2011,Ben-Jaffel2013,Llama2013}.
  
Furthermore, interactions of tidal and magnetic nature are likely to occur in 
systems with hot Jupiters \citep{Cuntz2000,Saar2004,Lanza2009,Lanza2010,Lanza2011,
Cohen2009,Cohen2010,Cohen2011}.
Stars can receive angular momentum from their hot Jupiters during the migration and 
circularization of the planet's orbit. The external input of angular momentum to the 
star reduces its decline with age during the main sequence due to wind losses.
Because of the connection between stellar activity, age and rotation, 
the stellar activity of the stars can be effectively boosted in presence of hot Jupiters,
mimicking a younger age as demonstrated for the systems of HD 189733 and Corot-2A 
\citep{Pillitteri2010,Pillitteri2011,Schroeter2011,Pillitteri2014,Poppenhaeger2014}. 
Tides can be produced on the stellar surface, with height proportional to $d^{-3}$  
so the closer is the planet the stronger is the effect, that occurs twice per orbital period.
However, in extreme cases, e.g. WASP-18 \citep{Pillitteri2014b}, 
the influence of the tidal perturbation could destroy the magnetic dynamo of stars with 
shallow convective zones. It is also suggested that the strong tidal stresses
in WASP-18 reduce significantly the mixing inside the convective zone, 
as evidenced by a high Li abundance. This is at odds with the finding that planet 
host stars have depleted more Li than single stars \citep{DelgadoMena2014,Bouvier2008,
Gonzalez2008,Israelian2004}.

Magnetospheric interactions between planetary and stellar magnetic fields can be the 
source of additional reconnection events, and perturbations of the magnetic field topology. 
As a result, more flares could manifest in systems with hot Jupiters than in single stars 
of the same age. Furthermore, the planetary magnetic field can efficiently trap the 
stellar wind and its angular momentum, modifying the losses of rotation and keeping the 
star rotating faster than expected for its age \citep{Cohen2011}.

\citet{Shkolnik03, Shkolnik05, Shkolnik08, Walker2008} 
reported evidences of chromospheric activity phased with the planetary orbital motion in 
the systems of $\mu$~And, HD~179949 and HD~189733, probed by Ca H\&K lines.  
Interestingly, all cases exhibit a phase shift: the site of enhanced activity 
on the stellar surface is $70-80\deg$ in the case of HD 179949  and $169\deg$ for 
$\mu$~And leading the sub-planetary point. 

In X-rays, we observed enhanced flare variability in HD~189733 after the eclipse 
of the planet ($\phi=0.5$) in a restricted range of phases 
$\phi=0.52-0.65$ \citep{Pillitteri2010, Pillitteri2011, Pillitteri2014}. 
The rate of such flare activity in \hd\ is higher than in stars a few Gyr old, and
similar to that of pre-Main Sequence stars or stars more active than the Sun.
Furthermore, this type of variability has not been observed at the planetary transits,
rising the question whether \hd\ has a prevalent X-ray activity at some orbital 
phases of its hot Jupiter \citep{Pillitteri2011,Pillitteri2014}.
An active spot on the stellar surface, at $\sim70-90\deg$ ahead of the sub-planetary point, 
magnetically connected with the planet, and co-moving with the planet had been hypothesized 
\citep{Pillitteri2014}. Such spot would emerge at the edge of the stellar disk when the 
planet is in a range of phases of $\phi=0.52-0.65$, originating thus a phased variability.

\citet{Lanza2010} explains this phenomenon with an analytical model of the stellar 
field. The main hypothesis is that a magnetic link between the star and the planet 
exists if the planet is sufficiently close to the star as in case of hot Jupiters. 
The lines of the magnetic field of the star could connect with the dipolar field of the 
planet via a bent path and generate thus the phase lag. 
The relative motion of the planet with respect to the star plays a key ingredient 
for generating such configuration. 
In \citet{Lanza2012} this model is further refined, with the planet motion inducing 
reconnection events near the surface of the star. 

\citet{Cohen2011} modeled the magnetic SPI in HD~189733 including a more realistic 
stellar magnetic field. Their magneto-hydrodynamic (MHD) simulations show 
that at certain phases the planet can 
induce reconnection events and flares if the planet encounter locally enhanced 
stellar  magnetic field.  The same model predicts the rate of exoplanet atmosphere loss, 
and the cometary tail that the planet motion is winding around its orbit.
\citet{Preusse2006} and \citet{Kopp2011} modeled the interaction between planet and star 
by means of alfv\`enic waves hitting the surface of the star, generated by the 
planet acting as a conductor moving within the stellar magnetic field and the stellar wind.  
This model can reproduce well the observed phase lag of chromospheric activity which 
arises from the relative motion of the planet and the star. 

Among the systems with hot Jupiters, \hd\ is a privileged target where 
to study SPI effects, planet evaporation and the dynamics of the planetary gas around the host star. 
It is composed of a K1.5V type star as primary component 
($P_{\rm rot} = 11.9$\,d, $d=19.3$\,pc from the Sun), and an M4 companion, \hd~B, at 3200 AU from 
the primary.  
\hd~A hosts a hot Jupiter class planet (HD~189733b) at a distance of only 0.031 AU  (about 8.5 stellar
radii) with an orbital period of $\sim 2.22$d \citep{Bouchy2005}.

In this paper we present observations of \hd\ obtained with the {\em Cosmic Origin Spectrograph} (COS)
on board HST. The aim of our observations is to obtain time resolved spectroscopy of \hd\ at the
post planetary eclipse phases, as observed with \xmm, in order to find signatures of SPI in 
FUV and to better recognize and understand any phased activity in \hd. 
Previous observations in FUV of \hd\ with HST COS and STIS have focused 
 at the planetary transits. These observation revealed the evaporation of the planetary atmosphere
and an estimate of the mass loss \citep{Lecavelier2010}, 
and asymmetry of ingress and egress implying a bow shock in front of the planet due to the 
interaction between the stellar wind and the planetary escaping atmosphere 
\citep{Lecavelier2012,Bourrier2013,Ben-Jaffel2013,Llama2013}.
Motivated by our findings in X-rays, we argue that observing \hd\ at the planetary eclipse
in FUV can offer new insights on the dynamics of the gas evaporating from the planet and of
the structure of the chromosphere and transition region of the star.
The paper is structured as follows: in Sect. \ref{analysis} we describe the observations
and the data analysis; Sect. \ref{results} details the results, in Sect. \ref{discussion} we discuss
the results, and in Sect. \ref{conclusions} we present our conclusions.

\section{Observations and data analysis} \label{analysis}
\begin{figure*}[t]
\includegraphics[width=0.5\textwidth]{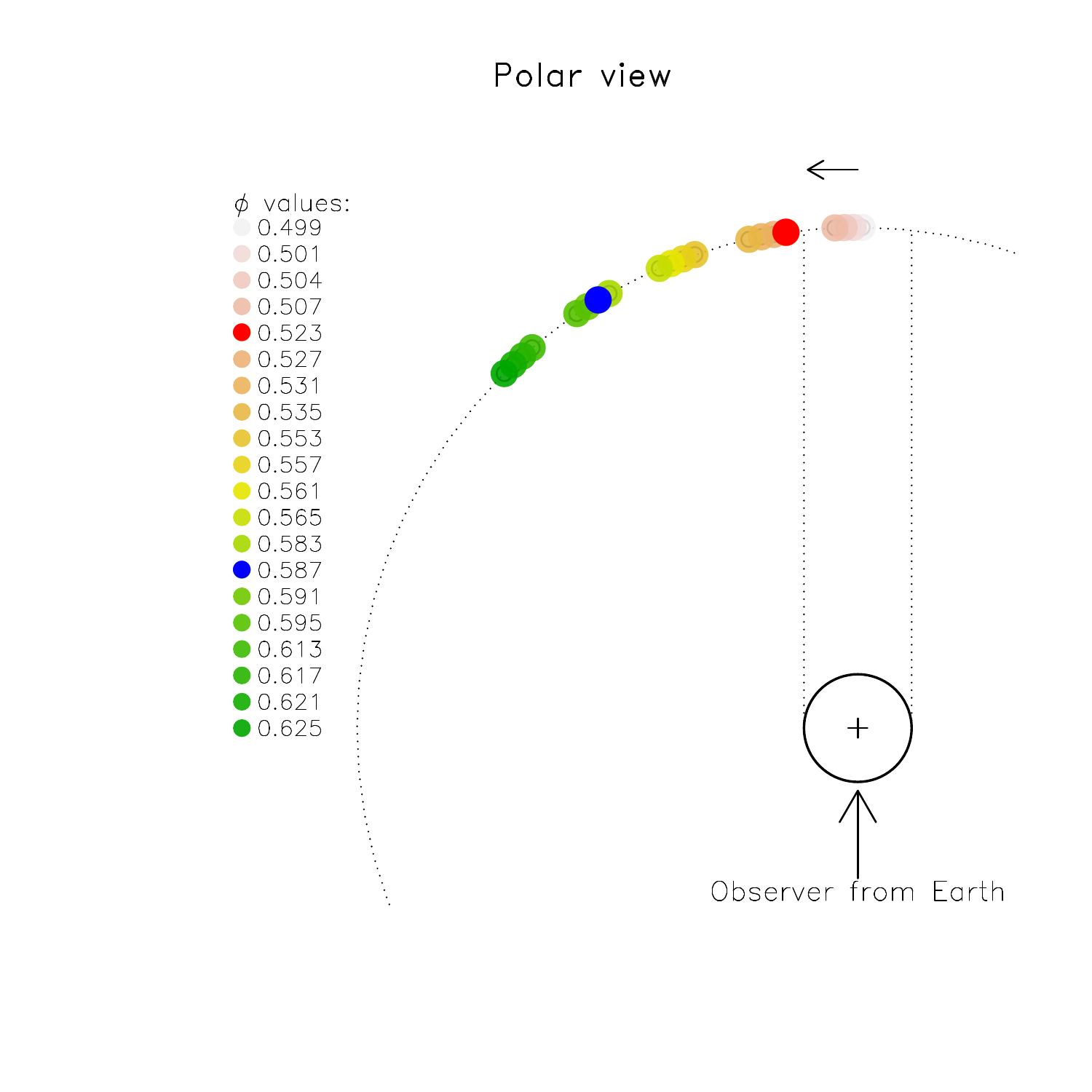}
\includegraphics[width=0.5\textwidth]{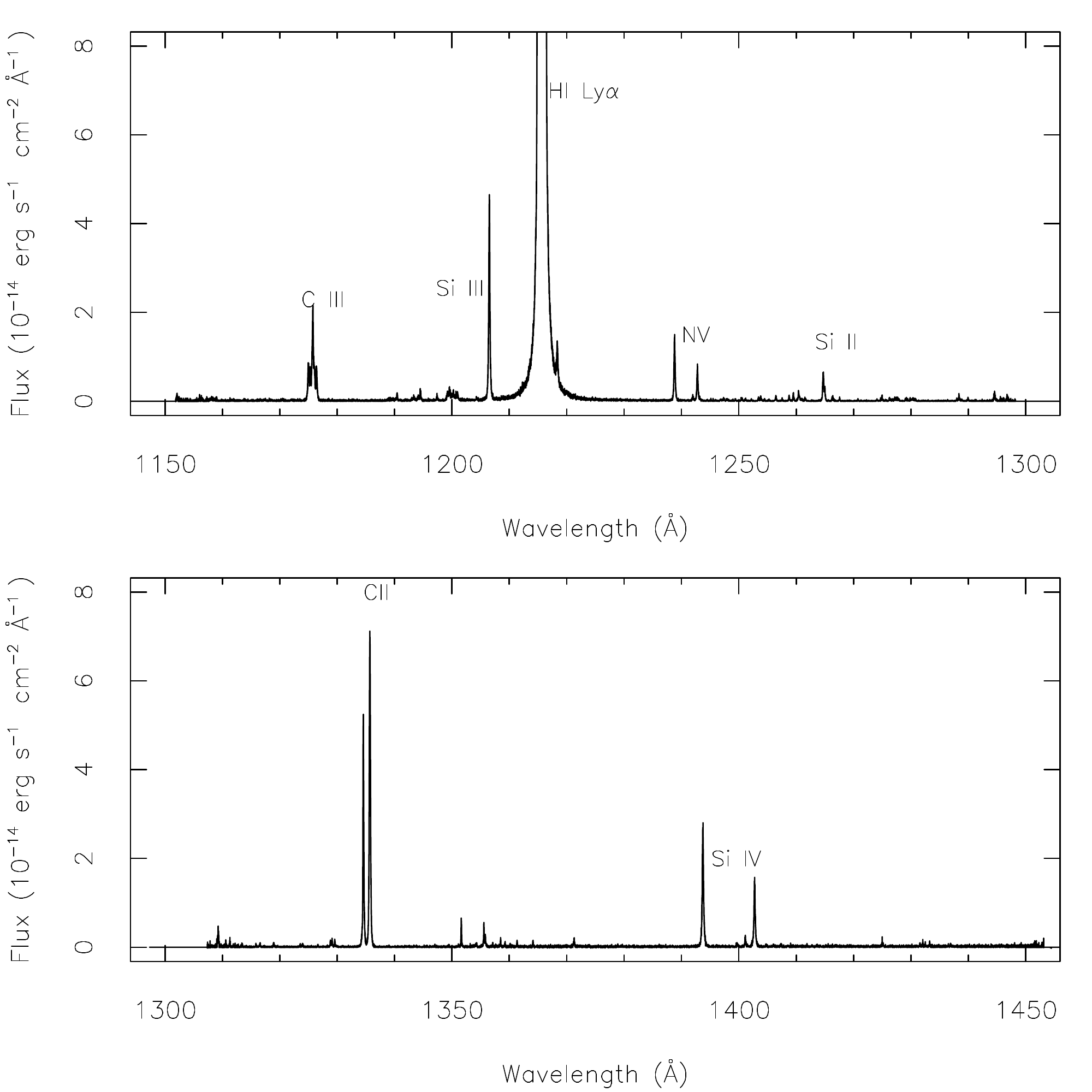}
\caption{\label{phases} Left panel: schematic of the planetary phases during the
COS exposures. Right panel: average spectrum obtained from the five HST orbits and the sum
of the COS exposures. We marked the main ion lines in the two wavelength ranges.
A detailed plot of the spectrum is available at \url{\atlasurl}. }
\end{figure*}

We observed \hd\ with HST and COS spectrograph on September 12 2013. The observations were
carried along five consecutive HST orbits, spanning the planetary orbital phases of 
0.4996 through 0.6258 (see Table \ref{logtab} and Fig. \ref{phases}, left panel). 
We used the grating G130M with central wavelength of 1300\AA\
that encompasses the range 1150\AA\ to 1450\AA\ and has a gap between the two detector
segments in the range $\sim1294-1309$ \AA.
The effective area of COS paired with grating G130M is above 1500 cm$^{2}$ with a peak 
of 2500 cm$^{2}$ around 1220\AA.   
For the transit and period, we used the ephemeris from \citet{Triaud2009}, which are 
based on the analysis of optical spectra obtained with the HARPS 
spectrograph at the ESO 3.6~m telescope in La Silla (Chile). 
When using the ephemeris from \citet{Agol2010},
based on Spitzer observations, this would result in a systematic shift of $\sim-249.1$~s 
with respect to the optical ephemeris of \citet{Triaud2009}.

\begin{figure}[t]
\includegraphics[width=0.5\textwidth]{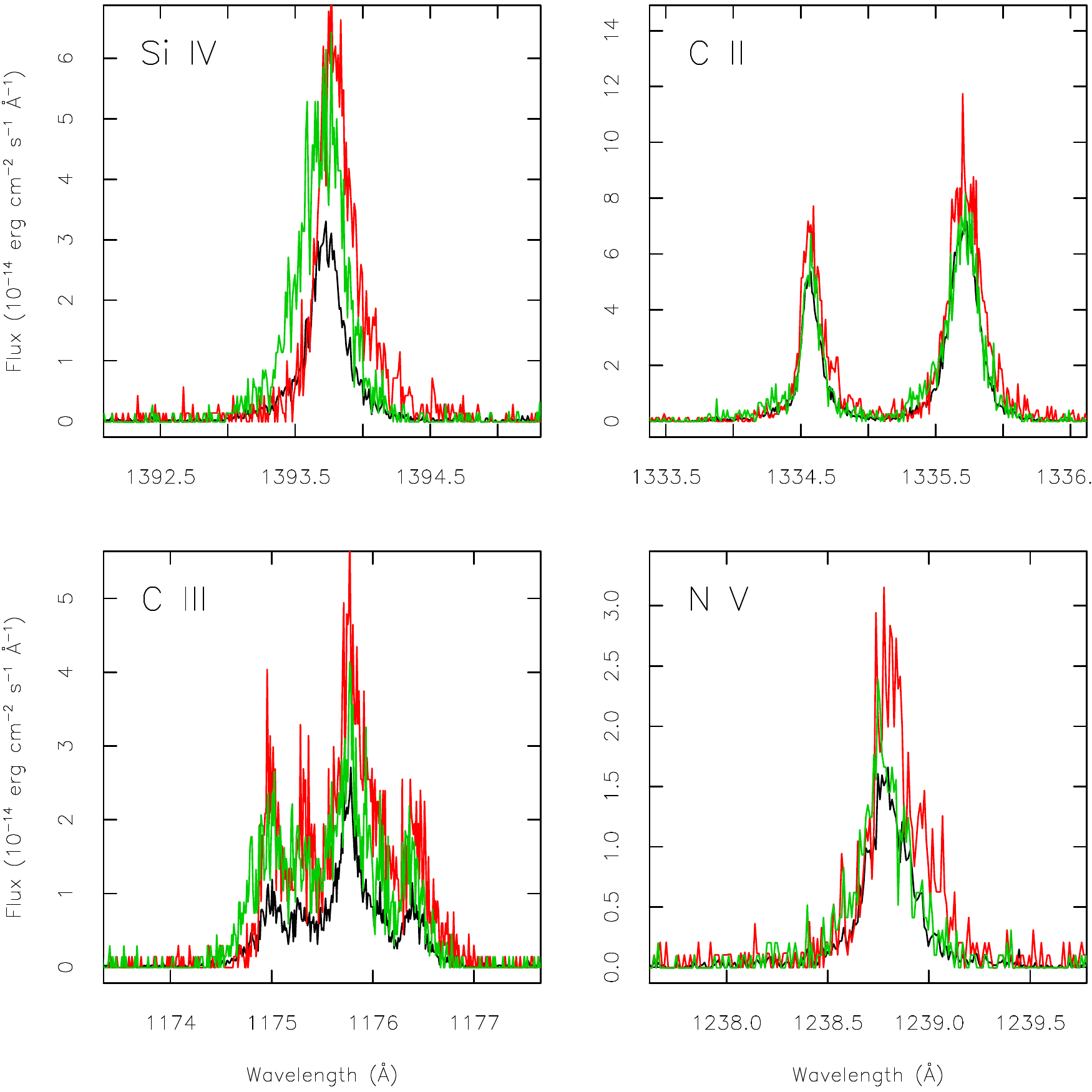}
\caption{\label{zoom_spec} Examples of line profiles in three spectra:
average of first orbit (black), exposure nr. 5 (red), and exposure nr. 14 (green).
The names of the ions are indicated in the panels.
The lines show an increase of flux and centroid shifts at these two exposures. }
\end{figure}

The average spectrum from the five orbits is shown in Fig. \ref{phases}, right panel.
The total exposure is about 12.1~ks.
The main lines in this range are HI Ly$\alpha$, Si II, Si III, and Si IV lines,
C II and C III lines, and N V lines. A list of these lines, with the peak of 
formation temperature, line intensity, and spectroscopic terms is given in 
Table \ref{linesummary}. These data are taken from {\sc Chianti} database 
\citep{CHIANTI,Landi2013}. However, other small lines are visible in our spectra
and missing from {\sc Chianti}. For identifying these lines we used
{\sc NIST} database \citep{NIST}. To help the identification of the lines, 
we produced an atlas of the spectrum\footnote{Available
at \url{\atlasurl}}, 
plotted in pieces of 10\AA. In each panel lines from {\sc Chianti} are labelled,
tickmarks at the NIST wavelengths of line of different elements 
(only those with relative accuracy $\le 50\%$) are indicated with different colors.  
Most of these lines are from Fe ions, Al, Si, and S. Also, airglow lines are 
marked with light blue bands across the spectrum.
Fig. \ref{zoom_spec} shows a close up plot of selected line profiles for three
spectra: average spectrum of first orbit, and exposures nr. 5 and nr. 14, that 
show two episodes of flux increases in several lines.

\begin{table*}
\begin{center}
\caption{\label{logtab} Log of the HST observations. }
\resizebox{0.8\textwidth}{!}{
\begin{tabular}{llccccc}
  \hline \hline
Orbit & Nr.& Exposure id. & Date & JD (days) & Planetary phase$^*$ & Exposure (s) \\ 
  \hline
I &   1 & lc0u01qjq & 2013:09:12T10:14:21 & 2456547.92663194 & 0.4996 & 431.0 \\ 
  &   2 & lc0u01qlq & 2013:09:12T10:23:27 & 2456547.93295139 & 0.5025 & 440.0 \\ 
  &  3 & lc0u01qnq & 2013:09:12T10:32:42 & 2456547.93937500 & 0.5054 & 440.0 \\ 
  &  4 & lc0u01qpq & 2013:09:12T10:41:57 & 2456547.94579861 & 0.5083 & 440.0 \\\hline 
II&  5 & lc0u01qrq & 2013:09:12T11:31:45 & 2456547.98038194 & 0.5239 & 640.2 \\ 
  &  6 & lc0u01qtq & 2013:09:12T11:44:20 & 2456547.98912037 & 0.5278 & 652.2 \\ 
  & 7 & lc0u01qvq & 2013:09:12T11:57:16 & 2456547.99810185 & 0.5318 & 654.2 \\ 
  & 8 & lc0u01qxq & 2013:09:12T12:10:14 & 2456548.00710648 & 0.5359 & 649.2 \\\hline 
III & 9 & lc0u01qzq & 2013:09:12T13:07:25 & 2456548.04681713 & 0.5538 & 648.2 \\ 
  & 10 & lc0u01r1q & 2013:09:12T13:20:17 & 2456548.05575231 & 0.5578 & 646.2 \\ 
  & 11 & lc0u01r3q & 2013:09:12T13:33:70 & 2456548.06466435 & 0.5618 & 646.2 \\ 
  & 12 & lc0u01r5q & 2013:09:12T13:45:57 & 2456548.07357639 & 0.5659 & 646.2 \\\hline 
IV& 13 & lc0u01r7q & 2013:09:12T14:43:70 & 2456548.11327546 & 0.5838 & 648.2 \\ 
  & 14 & lc0u01r9q & 2013:09:12T14:55:59 & 2456548.12221065 & 0.5878 & 646.2 \\ 
  & 15 & lc0u01rbq & 2013:09:12T15:08:49 & 2456548.13112269 & 0.5918 & 646.2 \\ 
  & 16 & lc0u01rdq & 2013:09:12T15:21:39 & 2456548.14003472 & 0.5958 & 646.2 \\\hline 
V & 17 & lc0u01rfq & 2013:09:12T16:18:50 & 2456548.17974537 & 0.6137 & 648.2 \\ 
  & 18 & lc0u01rhq & 2013:09:12T16:31:42 & 2456548.18868056 & 0.6177 & 646.2 \\ 
  & 19 & lc0u01rjq & 2013:09:12T16:44:32 & 2456548.19759259 & 0.6218 & 646.2 \\ 
  & 20 & lc0u01rlq & 2013:09:12T16:57:22 & 2456548.20650463 & 0.6258 & 646.2 \\\hline
\end{tabular}
}
\end{center}
Note: orbits are labelled with I, II, III, IV and V roman numbers.
Four dithered exposures are obtained for each orbit, thus a total of 20 exposures are obtained.
Exposure times are of $\sim640-650$ s during all orbits but I, where exposures were $430-440$ s.
Ephemeris for calculating planetary phases are from \citet{Triaud2009}.
\end{table*}

We label the orbits with I, II, III, IV and V roman numbers.
For each orbit we have available four dithered exposures as part of the ordinary strategy of 
data acquisition.
The dithering along the dispersion axis reduces the effects of inhomogeneous sensitivity pixel by
pixel. The typical exposure of the single exposures are about 640$-$650 s. Only during the first
orbit the exposures were shorter, of the order of $430-440$ s, because of the initial target 
acquisition overhead. 

With {\sc IRAF}  and {\sc CALCOS} package we have transformed the calibrated tables and spectra
in a suitable format to be read from other software (e.g., Python and R).
Our analysis is aimed  at a fine time resolved spectroscopy, by exploiting the capability
of COS to acquire single photons with their arrival time. This feature allows us to further
slice the exposures obtained during the five orbits of our observation. In particular,
two of the 20 exposures, corresponding to the 5th and 14th ones, were further splitted into three time
intervals of $\sim200$ s each in order to follow in greater details the change of line profiles and fluxes
as described in the results session. To this purpose we used the tasks {\sc splittag} and {\sc calcos}
to split the time tagged tables of selected photons and accumulate the calibrated spectra relative 
to each interval.  

For the most prominent lines in each spectrum, we have measured fluxes, line FWHMs and centroid 
with respect to the rest wavelength of the lines with an R script. The script calculates the flux as the 
integral of the line counts in a range of $\pm1$\AA\  around the rest wavelength of the line.
For C II lines at $1334-1335$\AA\ we used 0.5 \AA\ window to avoid cross contamination from the nearby
line. Given the spectral type of the star, the background is consistent with zero for all the lines. 
The centroid is calculated as the intensity weighted mean over the range of wavelength. 
The {\em Full Widths at Half Maximum} (FWHMs) are calculated as the wavelength intervals where
the flux is higher than half of the peak of the line.  

\begin{table}[t]
\centering
\caption{\label{linesummary} 
List of prominent lines in the spectral range 1150\AA\--1450\AA\ 
used in our COS observations. Lines falling in the gap 1294-1309\AA\ are not listed.} 
\resizebox{0.6\textwidth}{!}{
\begin{tabular}{lccc}
  \hline
\hline
Ion & Wavelength &  $\log(\mathrm T)$ & Transition \\ 
   &  \AA\ & K &  \\ \hline
  C III       & 1174.93 &  4.80 &  $2s 2p ^3P_1 - 2p^2\ ^3P_2$ \\
  C III       & 1175.26 &  4.80 & $2s 2p ^3P_0 - 2p^2\ ^3P_1$ \\
  C III       & 1175.59 &  4.80 & $2s 2p ^3P_1 - 2p^2\ 3^P_1$ \\
  C III       & 1175.71 &  4.80 & $2s 2p ^3P_2 - 2p^2\ 3^P_2$ \\
  C III       & 1175.99 &  4.80 & $2s 2p ^3P_1 - 2p^2\ 3^P_0$ \\
  C III       & 1176.37 &  4.80 & $2s 2p ^3P_2 - 2p^2\ 3^P_1$ \\ 
  S III       & 1200.96 &  4.70 & $3s^2 3p^2\ 3^P_2 - 3s 3p^3\ 3^D_3$ \\
  Si III      & 1206.50 &  4.70 & $3s^2  1^S_0 - 3s 3p ^1P_1$ \\ 
  Si III      & 1206.56 &  4.80 & $3s 3p ^1P_1 - 3s 3d ^1D_2$ \\
  HI          & 1215.67 &  4.50 & $1s^2 S_{1/2} - 2p^2 P_{3/2}$ \\
  HI          & 1215.68 &  4.50 & $1s^2 S_{1/2} - 2p^2 P_{1/2}$ \\
  OV          & 1218.34 &  5.30 & $2s^2\ ^1S_0  -  2s^2 p  ^3P_1$ \\
  NV          & 1238.82 &  5.20 & $1s^2\ ^2s_2 S_{1/2} - 1s^2 2p ^2P{3/2}$ \\
  NV          & 1242.81 &  5.20 & $1s^2 2s ^2S_{1/2} - 1s^2 2p  ^2P_{1/2}$ \\
  Si II       & 1264.74 &  4.50 & $3s^2 3p ^2P_{3/2} - 3s^2 3d  ^2D_{5/2}$ \\
  C II        & 1334.54 &  4.50 & $2s^2 2p ^2P_{1/2} - 2s 2p^2\ ^2D_{3/2}$ \\
  C II        & 1335.66 &  4.50 & $2s^2 2p ^2P_{3/2} - 2s 2p^2\ ^2D_{3/2}$ \\
  C II        & 1335.71 &  4.50 & $2s^2 2p ^2P_{3/2} - 2s 2p^2\ ^2D_{5/2}$ \\
  Si IV       & 1393.76 &  4.90 & $3s      ^2S_{1/2} - 3p       ^2P_{3/2}$ \\
  O IV        & 1401.16 &  5.10 & $2s^2 2p ^2P_{3/2} - 2s 2p^2\ ^4P_{3/2}$ \\
  Si IV       & 1402.77 &  4.90 & $3s^2 S_{1/2}      - 3p       ^2P_{1/2}$ \\ \hline
  \end{tabular}
}
\end{table}

\section{Results} \label{results}
The average spectrum of the five HST orbits is presented in Fig. \ref{phases}, right panel.
We can distinguish several lines of Si, C, N, and O ions, 
other than the H I Ly$\alpha$ line, which is the main feature. 
Also the individual spectra of the 20 dithered exposures are of excellent quality and allow firm
measurements of line fluxes and centroids. 
Plots of the line doublets of Si IV, N V, C II, and the multiplet of C III at 1175\AA\ at the
different orbits are shown in Fig. \ref{zoom_spec}.
Most of the results are based on the analysis of Si IV doublet at 1393.7/1402.5 \AA, 
N V doublet at 1238.8/1242.8 \AA, Si II line at 1264.7, Si III blend at 1206.5.
Tables \ref{lineflu1}, \ref{lineflu2} and \ref{lineflu3} lists the fluxes, the centroids and FWHMs 
of the lines measured in each of the twenty exposures acquired in the five orbits.
Figs.~\ref{fluxes}, \ref{fluxes_archive} and \ref{centroids} show the fluxes, the
line centroids, and FWHMs as a function of the planetary orbital phases. 
The fluxes of the lines shows two main increases or brightenings during exposures nr. 5 and 14, 
seen in Si and C ion lines, as well as in N~V. During the rest of the 
observations small variability is detected at level of $1-2\sigma$, similarly
to what is observed at the planetary transit phase using archival observations 
(cf. Fig. \ref{fluxes_archive}).
We discuss the quiescent spectrum and the two brightenings in the following sections.  

\subsection{Quiescent spectrum}
Data from the first HST orbit were taken during planetary eclipse, and should be devoid of 
any direct planetary contribution.  
The line fluxes from the first orbit can be assumed as the basal fluxes from the star alone,
given that the planet is obscured by its star. 
During orbits III and V the line fluxes are similar to the basal flux, 
suggesting that any component of planetary origin is negligible at these epochs.
We have used archival data of \hd, obtained at the planetary transit, with COS 
and the same grating (4 HST orbits, exposures of $900$s each), to check the extent of 
line variability at those epochs. 
Variability of line fluxes of Si and C is recognized in the archival
spectra at 2-3 $\sigma$ level (see Fig. \ref{fluxes_archive}).
Overall, during HST orbits I, III and V, the line fluxes are consistent to
within 1$\sigma$, analogously to the transit observations.
Moreover, during a planetary transit, the lines of C II doublet and Si III exhibited a rapid increase
and decay at level of $2\sigma$, markedly visible in the C II line blend at 1335.7 \AA. 
A slow trend superimposed to the rapid increase is also evident in the same C II lines.

The ratio of line intensities gives a diagnostic of plasma temperature, 
determining the basic thermal structure of the emitting plasma and inferring
its time evolution.  
We plot in Fig. \ref{si_ratio} (top row) the ratios of fluxes as a function of the 
planetary phase for three Si lines: Si IV \@ 1402.77\AA, Si III \@ 1206.50 \AA, and 
Si~II \@ 1264.74. We have assumed as basal flux the average of values obtained in orbit I,
in exposures $1-4$, when the planet is hidden behind the star. 
For most of the observations (namely, during orbits I, III and V, and part of orbits II and IV),
the fluxes and the line ratios show a small scatter within $1-2\sigma$, similarly
to what observed at the planetary transit phase. The three ratios Si III / Si IV, Si II / Si IV
and Si II/ Si III probe three different temperature regimes. 
The evolution of the temperature derived from the ratios of 
Si lines is plotted in Fig. \ref{si_ratio} (bottom panels). 
The quiescent plasma has three components roughly at 80,000 K, 50,000 K, and 25,000 K. 

The ratio of lines in doublets of Si IV, N V and C II are sensitive to opacity effects
\citep{Mathioudakis1999,Bloomfield2002}. In optically thin plasma, the ratio of the pairs of lines is around
2 for the Si IV and N V, and $\sim1.8$ for C II. Any departure from these values is 
suggesting that dense plasma is present. We have obtained the ratios for Si IV, N V and C II
doublets for each exposure, and compared these with {\sc Chianti} predictions (Fig. \ref{opacity_ratio}). 
We do not observe strong departures from the expected values (the scatter is within $1\sigma$
for most of the exposures).  
On average, N~V ratios show systematic lower values and C~II ratios show 
an excess with respect to the theoretical value, while Si~IV ratios are in better agreement with 
the expected  value of 2. No clear departure is found at the two brightenings. Overall, 
we conclude that the conditions for optically thin plasma are met during exposures, and the small
systematic difference in N~V and Si~IV could be due to uncertainties in the atomic database.  

\subsection{Line variability}
Two main episodes of flux variability are observed at phases $\phi\sim0.525$ and  $\phi\sim0.59$,
during exposures 5 and 14. 
We followed the evolution of the spectrum of \hd\ during theses two exposures in more detail, 
accumulating three spectra from the photons recorded 
in three shorter time intervals of length $\sim200$s, as described in Sect. \ref{analysis}. 
In Tables \ref{lineflu1}--\ref{lineflu3} we report the measurements obtained in 
these six time sub intervals of $\sim200$s each (labelled with 5a, 5b, 5c, and 14a, 14b, 14c,
respectively).

The line variability is significant at a level of $8-10 \sigma$.
The difference of variability at planetary transits and post eclipse phases
is reminiscent of the different X-ray variability observed at transit 
and post eclipse phases  \citet{Pillitteri2014,Pillitteri2011}. This suggests the remarkable 
behavior of the transition region and the corona of  \hd\ after the planetary eclipse, 
 and point to a major role of its hot Jupiter.

We discuss in more details each episode of variability in the next two sub-sections.
\begin{figure*}
\includegraphics[width=0.51\textwidth]{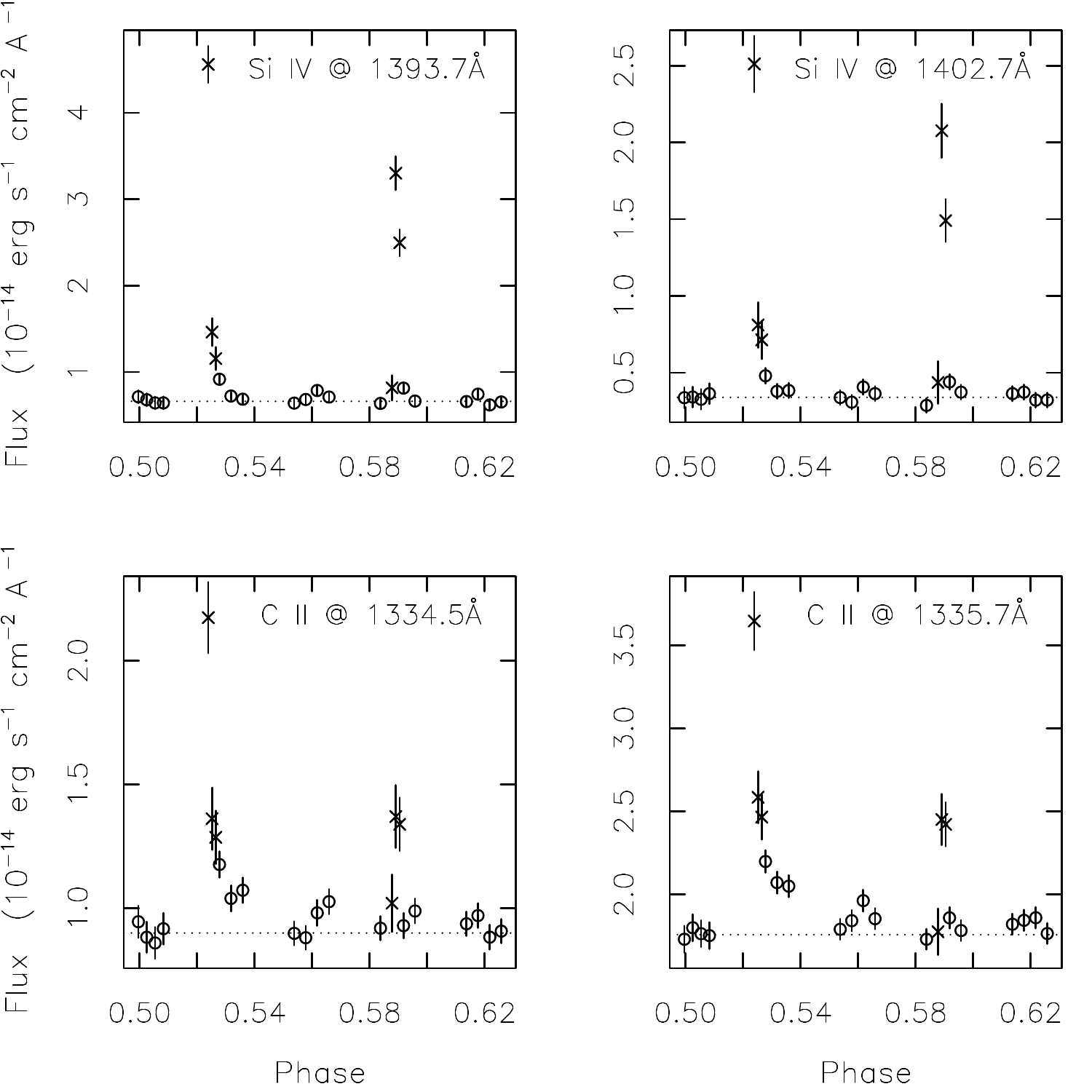}
\includegraphics[width=0.51\textwidth]{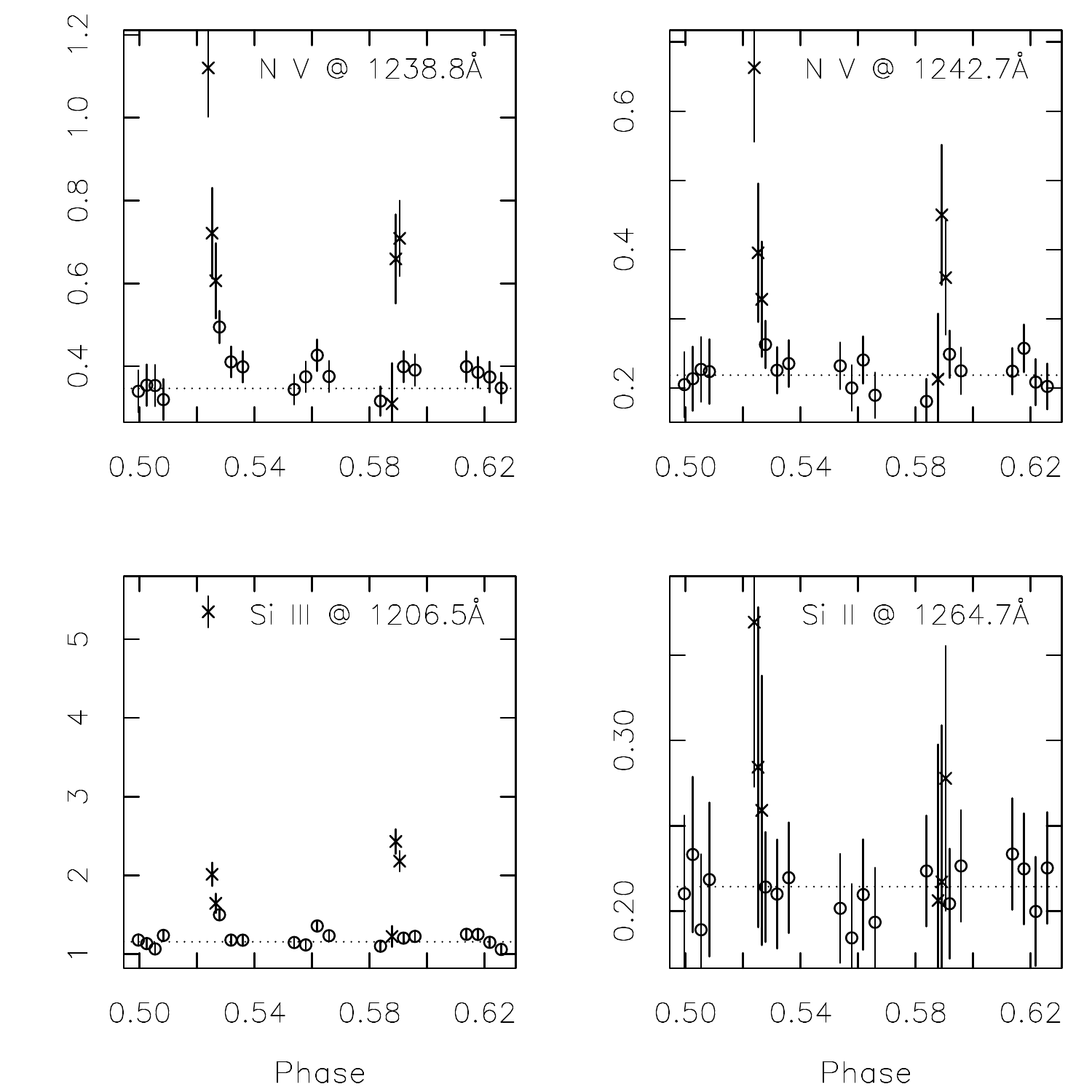}
\caption{\label{fluxes} Fluxes of lines as a function of the
planetary phases from spectra obtained during our program. 
The names and the wavelengths of the lines are indicated in each panel. The horizontal
line represents the average stellar flux obtained from the first four spectra, i.e.,
when the planet is completely obscured by the star. Two main episodes of variability
are observed at phases 0.525 and 0.588 with duration $t\le400$ s, for which we 
obtained sub-interval spectra (cross symbols).}
\end{figure*}

\begin{figure*}
\includegraphics[width=0.5\textwidth]{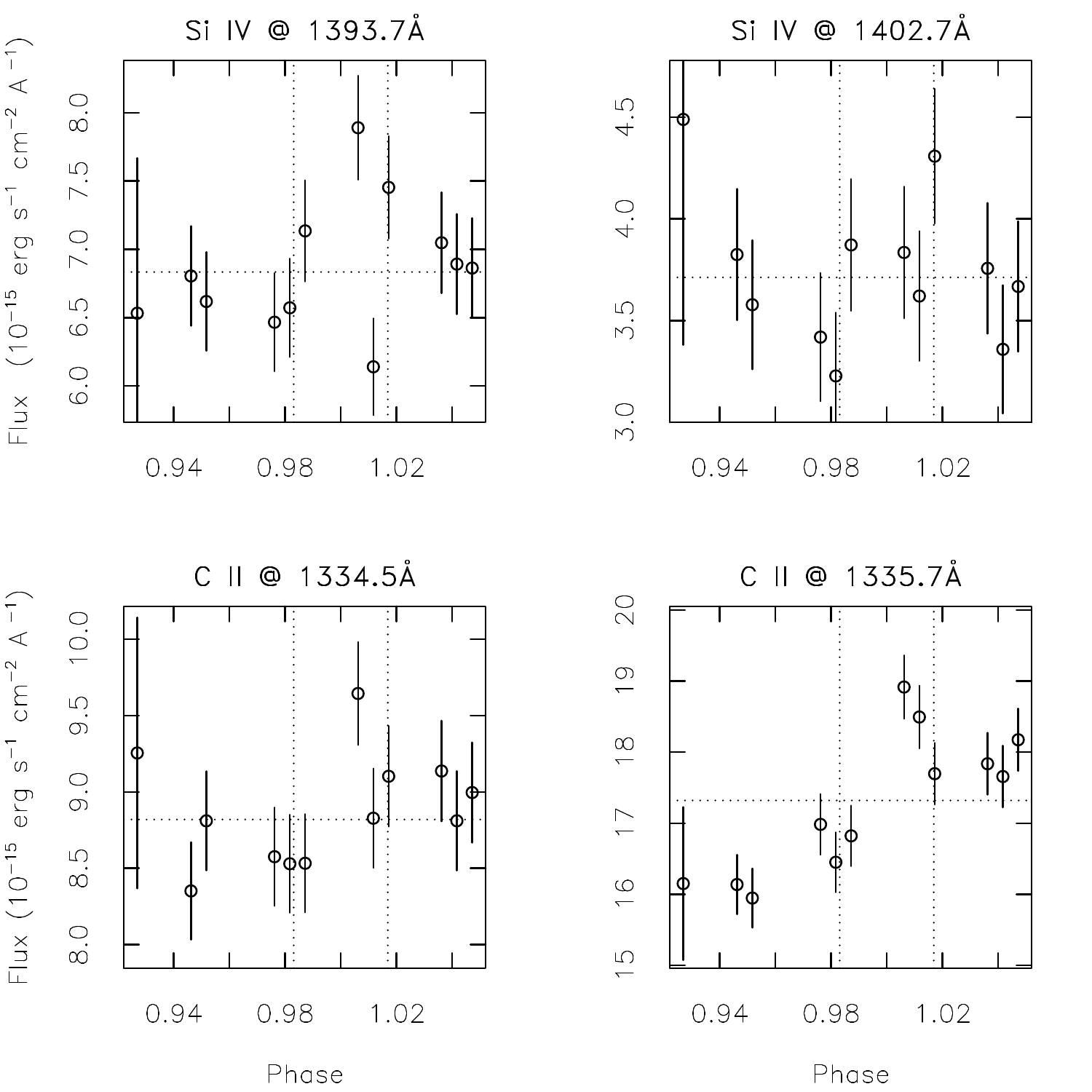}
\includegraphics[width=0.5\textwidth]{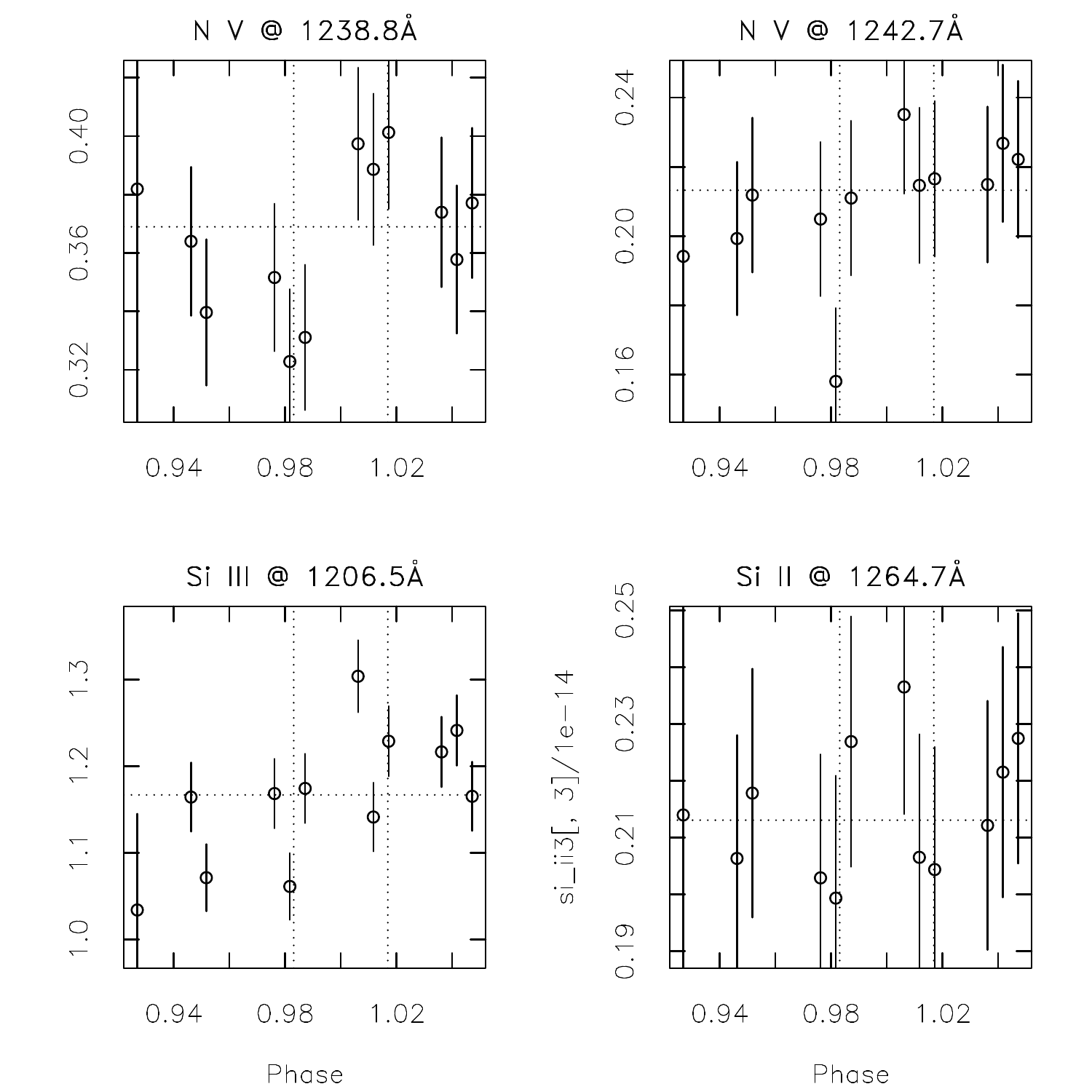}
\caption{\label{fluxes_archive} As in Fig. \ref{fluxes}, fluxes of lines as a function of the
planetary phases from archival spectra obtained at the planetary transit. 
The names of lines are indicated on the top axis. The horizontal
line represents the average stellar flux. Variability is observed at $\le3\sigma$ level
only.}
\end{figure*}

\subsection{First flux increase} 
The line profiles of Si and C ions during time intervals 5a, 5b, 5c (top row), 
and 14a, 14b, 14c (bottom row) are plotted in Appendix (Figs. \ref{si_iv_profiles} through 
\ref{n_v_profiles2}).
We marked the centroid and the nominal positions of the lines.
The figures shows in details the changes of flux, centroid and widths of these lines
on a time scale of $\sim200$ s. 
 
\begin{figure*}
\includegraphics[width=\textwidth]{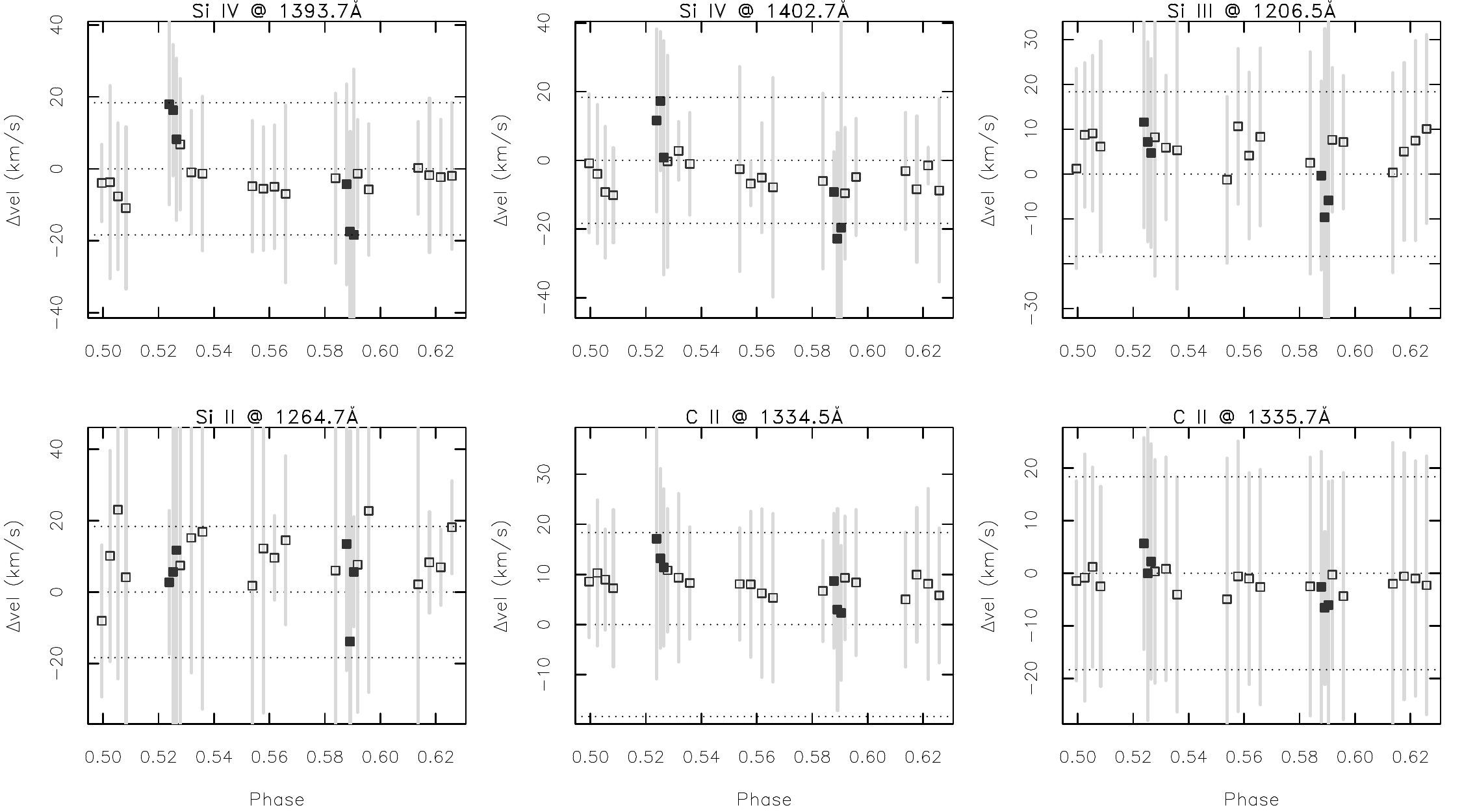}
\caption{\label{centroids} Centroids and FWHMs (vertical segments) of lines as a function of
the planetary phases. The horizontal lines mark the values of $\pm18.3$ km/s, i.e., the 
velocity of the sub-planetary point at the stellar surface co-moving with the planet.
For Si ions the centroid shifts at phases 0.525 and 0.59 are quite close to these values.
Filled symbols are the points from sub-interval spectra.}
\end{figure*}

At the beginning of the exposure 5 (second orbit), the line fluxes are already elevated. 
The interval 5a is the first temporal segment in which we observed \hd\ 
when the planet is appearing on view after its eclipse (see the schematic on 
Fig. \ref{phases}, left panel, phase range = $0.523-0.527$). 
We were unable to observe the phases between 3rd and 4th contact, when the planet 
emerged after the eclipse, so the increase of the line fluxes could have started 
while the planet was emerging. The line fluxes decrease on a time scale of 400 s
restoring to the basal level for the rest of the HST orbit and the subsequent one, 
until the second flux increase at phase $\sim0.59$. 
Small variability is seen during the third HST orbit (Fig. \ref{fluxes}), 
at a level of 2$\sigma$ variations, and of the same level of variability observed at 
the planetary transits.

Fig. \ref{centroids} shows the line centroid shifts with respect to the
line wavelength at rest (in km/s) and the FWHMs, as a function of the planetary phases.
The uncertainties of centroids are estimated to be  $\sim4-5$ km/s.
In the same plots, the horizontal lines correspond to the velocity of a point on the
stellar surface co-moving with the planet orbital period. We also recall that the 
stellar period is comprised between 11.9d and 16d \citep{Fares2010}, corresponding to the velocities
$2.5-3.4$ km/s and thus smaller than the speed of the point co-moving with the planet (18.3 km/s). 

During the first increase of flux, the centroids of all ions but Si II at the peak
of the brightening are shifted red-ward. For Si IV the pattern is the clearest, with
velocity shifts consistent with the co-moving planetary speed. 
For C~II line at 1334.5 \AA\ the centroid of the line is systematically shifted toward
$\sim+9$ km/s, and on top of this bias we observe the same pattern of shifts as in the case
of Si IV lines. This systematic redshift is also seen in the archival spectra taken
at the planetary transit. A more accurate de-blending analysis with IRAF shows that
the lines of C~II can be fitted with two Vogt profiles, one consistent with the RV shifted 
stellar wavelength, and the other profile red-shifted by $\sim20-23$ km/s which is still
consistent with the planetary co-moving speed (18.3 km/s), given the uncertainties of
centroid estimates. 

Similar results are obtained when using a cross-correlation technique with {\sc fxcor} task 
in IRAF. First, we subtracted spectra of exposures 5a, 5b, 5c, and 14a, 14b, 14c 
from the average spectrum  of the orbit I 
(after rebinning them a factor 4x of the original resolution).
For each residual spectrum, a spectral portion around the C II and Si IV doublets was
cross-correlated with the average spectrum of orbit I. We find that in exposures 
5a, 5b, 5c the residual spectrum is red-shifted by up to $20\pm5$ km/s, while in exposures
14a, 14b, 14c the residual spectra are blue-shifted of $-20\pm5$ km/s with respect to the
average, basal spectrum of orbit I.

\subsection{Second flux increase \label{secondevent}} 
The second event of variability started at phase $\sim 0.588$ and lasted for about 400 s.
It is weaker than the first event. Remarkably, this event along with the X-ray flares observed in
2009, 2011 and 2012 \citep{Pillitteri2010,Pillitteri2011, Pillitteri2014} occurred
in a very restricted range of phases ($\phi \sim0.52-0.6$).
Fig. \ref{centroids} shows that the lines and in particular 
Si~IV and Si~III during this brightening exhibit a remarkable blue-shift, with Si~IV values 
similar to the planetary co-moving speed ($\sim-18$ km/s). 
The interpretation of this blue-shift is that hot material on the surface of the
star, perhaps a hot spot at temperatures of $\sim10^5 K$, is co-moving with the planetary motion
and emerging at the stellar limb.

In Fig. \ref{ecdf} we plot the empirical cumulative distribution functions (ECDFs) of the time-tagged
events that compose three lines of different temperatures of formation, 
namely C II at 1334.5\AA, Si IV at 1393.7\AA, and N V at 1238.8\AA. 
Fully exploiting the time tagging feature of COS,  these curves allow us to evidence 
the timing of the line increases.
The two sets of distributions are created for events accumulated within 0.5\AA\ from the nominal line 
center, during exposures 5 and 14 (HST orbits II and IV) that host the two brightenings. 
During orbit II, the spectral lines start to brighten  simultaneously, as
demonstrated by the deviation from the purely constant rate (the linear ramp in dotted line). 
However, Si IV  increases its flux faster than N V and C II lines during the first brightening. 
The second brightening shows different ECDF shapes: there is a delay between the onset of 
brightening in N V, Si IV and C II, with a peak of N V growth is at phase 0.588. 
N~V line brightening starts earlier and at a faster rate than Si~IV line.
Here C II line starts last and it is the slowest, nearly exactly the opposite behavior
than in the first brightening. 

The nature of the two brightenings appears different: the first is a gradual event,
almost simultaneous in the three lines, and less impulsive than the second one.  
The bulk of plasma temperature in this event is near the peak temperature  of Si IV line, 
while relatively less plasma is present in the temperature range of the other two lines.   
The second event is an impulse, with hotter plasma appearing first (T$\ge$100,000 K), then in a 
sequence intermediate temperature plasma appears (60,000 K) and, eventually, plasma as cool as 30,000 K.
The nature of this second event is plausible with a flaring episode due to an accreting event
with a  subsequent cooling of the plasma to return to the pre-flare thermal conditions.
The analysis of the differential emission measure in the next sub-section helps to better
understand the thermal structure of the plasma during quiescence and the two brightenings.

\begin{figure*}
\includegraphics[width=0.9\textwidth]{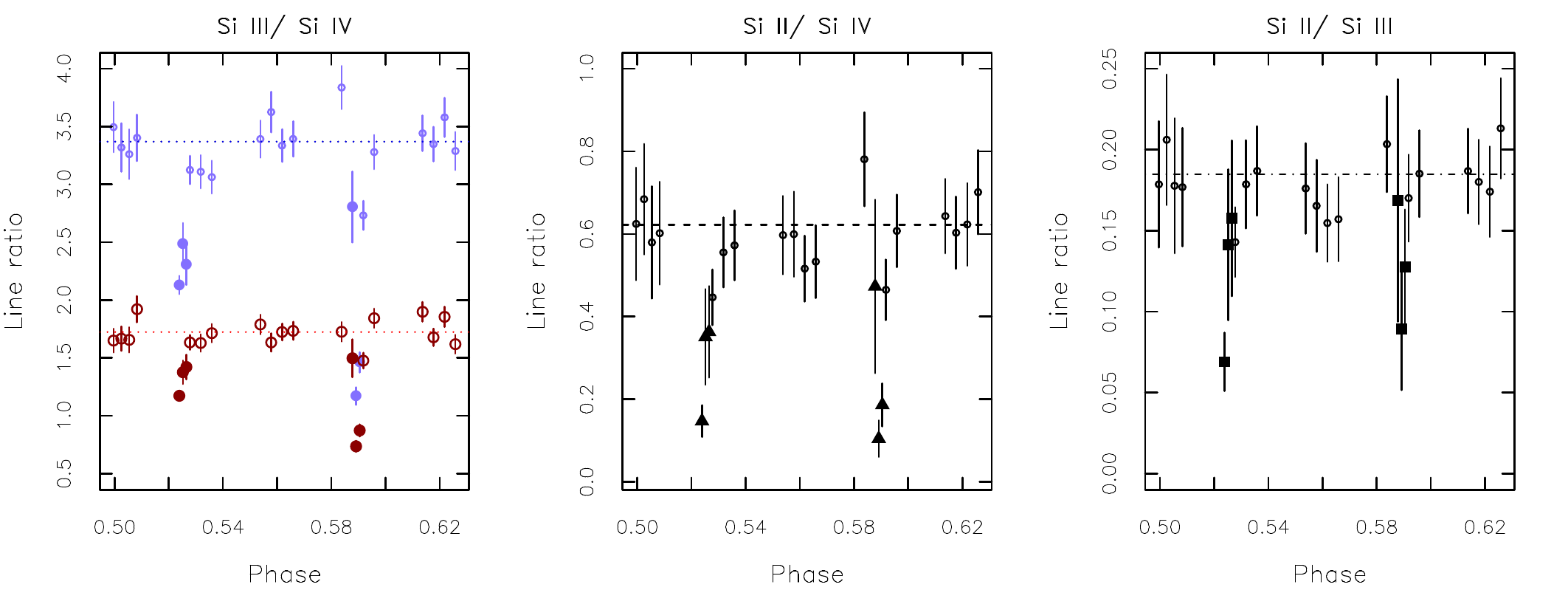}\\
\includegraphics[width=0.9\textwidth]{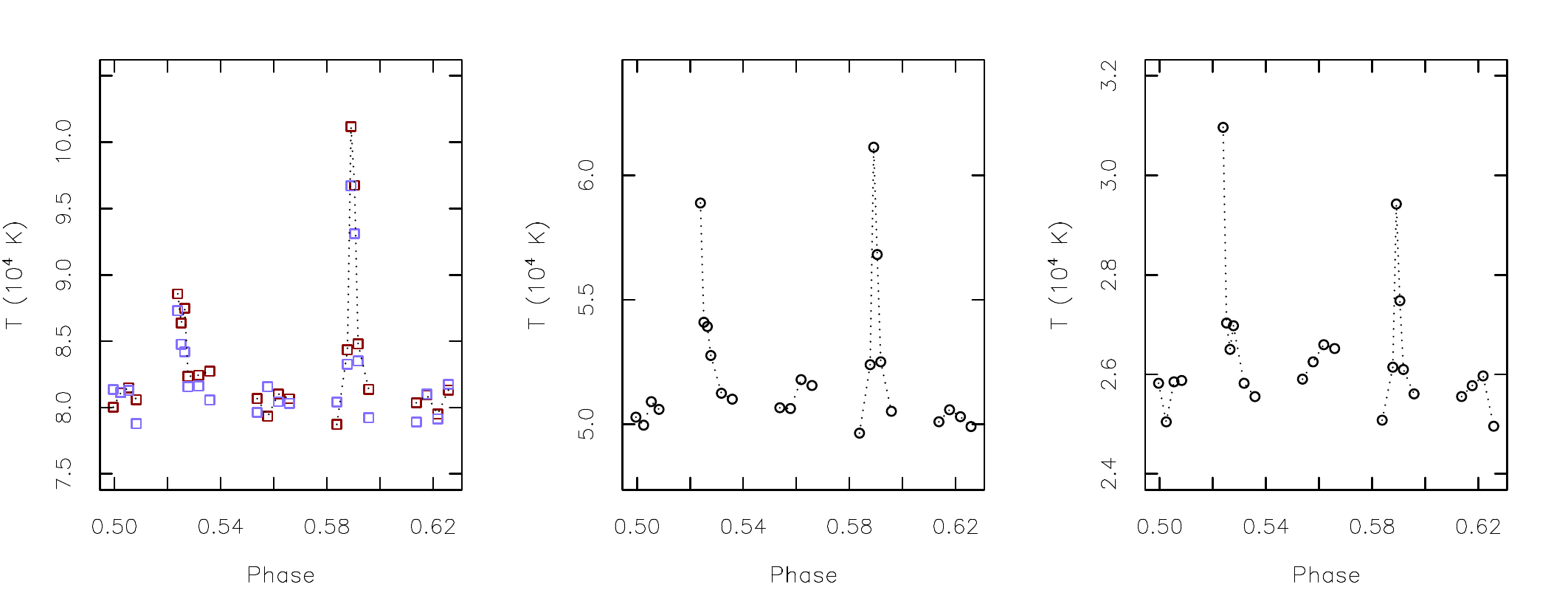}
\caption{\label{si_ratio}Top row: 
ratio of Si lines as a function of the planetary orbital phase. 
Filled symbols are the points from sub-interval spectra.
Left panel: Si III / Si IV \@ 1402.5\AA (blue symbols), 
Si III / Si IV \@ 1393.7\AA (red symbols). Central panel: 
Si II / Si IV \@ 1402.5\AA. Right panel: Si II / Si III.
Bottom row: temperature vs. planetary phase derived from line ratios. 
The comparison of the two peaks shows that the second brightening has been 
hotter in temperature than the first one.
}
\end{figure*}

\begin{figure*}
\includegraphics[width=0.9\textwidth]{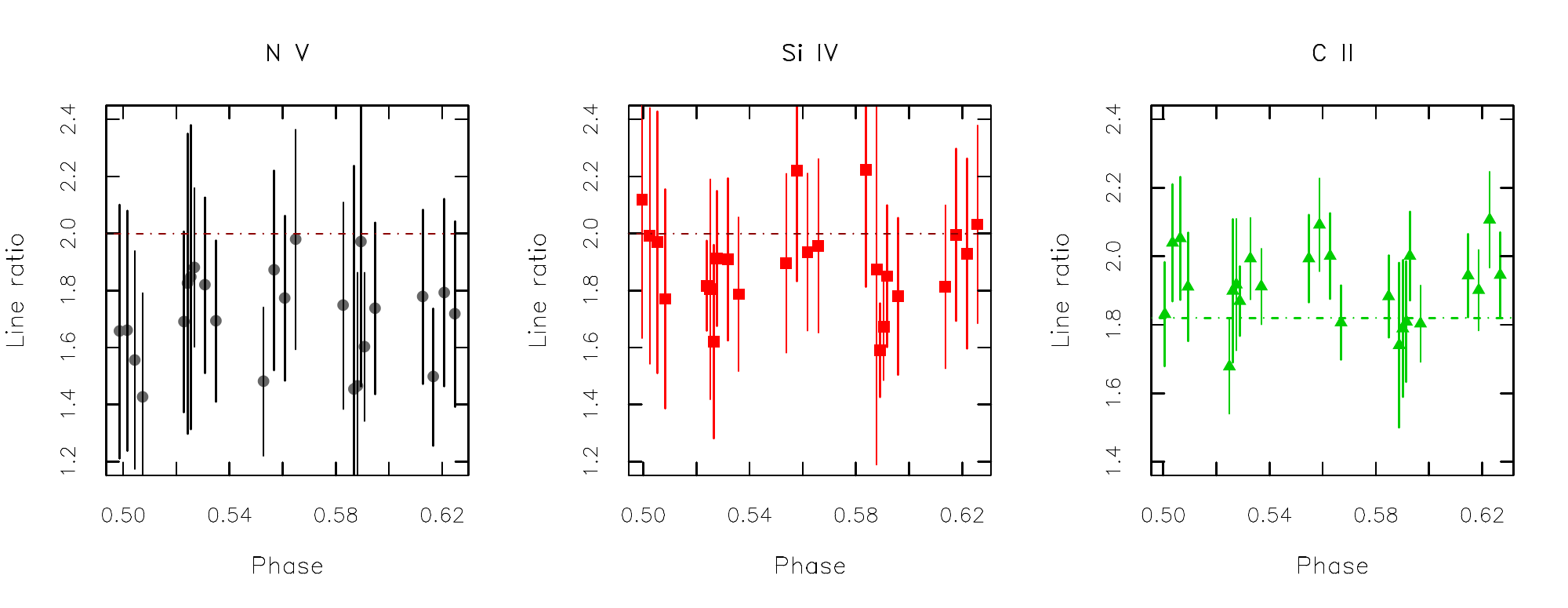}
\caption{\label{opacity_ratio} Ratio of lines of doublets of N~V, Si~IV and C~II.
The horizontal lines mark the expected values in optically thin plasma. Opacity effects
are not detected. }
\end{figure*}

\begin{figure*}
\begin{center}
\includegraphics[width=0.7\textwidth]{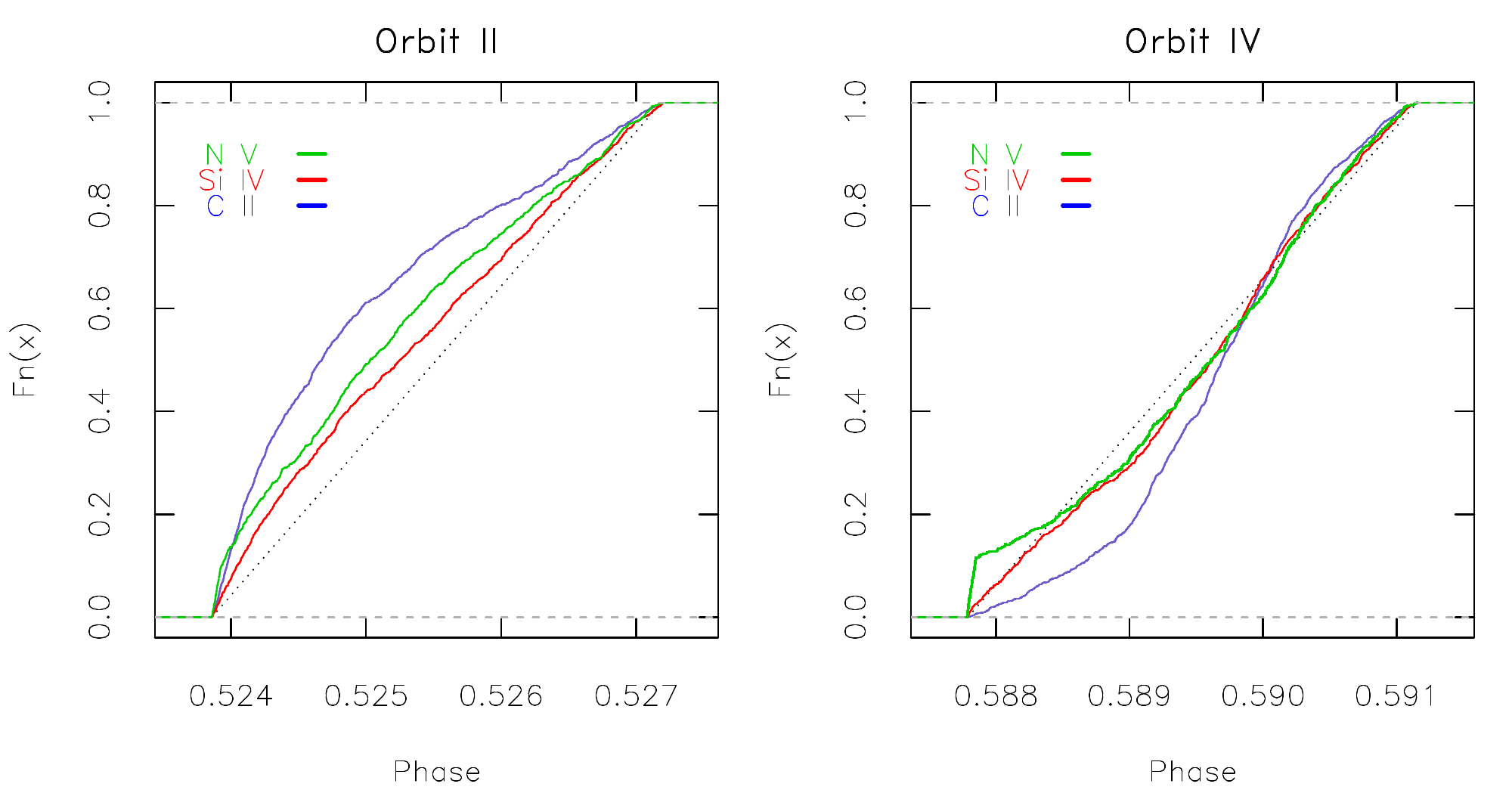}
\end{center}
\caption{\label{ecdf} Empirical cumulative distribution functions of events of 
lines of C II 1334.5 \AA\ (blue), Si IV 1393.7\AA\ (red), and N V 1238.7\AA\ (green).
The curves refer to the orbits II (left panel) and IV (right panel) during the two
line brightenings. A different rise of the lines is observed during the two brightenings,
slowly rising the first one, hotter rapid and impulsive the second one.}
\end{figure*}

\subsection{Differential emission measure}
\begin{figure*}
\begin{center}
\includegraphics[width=0.99\textwidth]{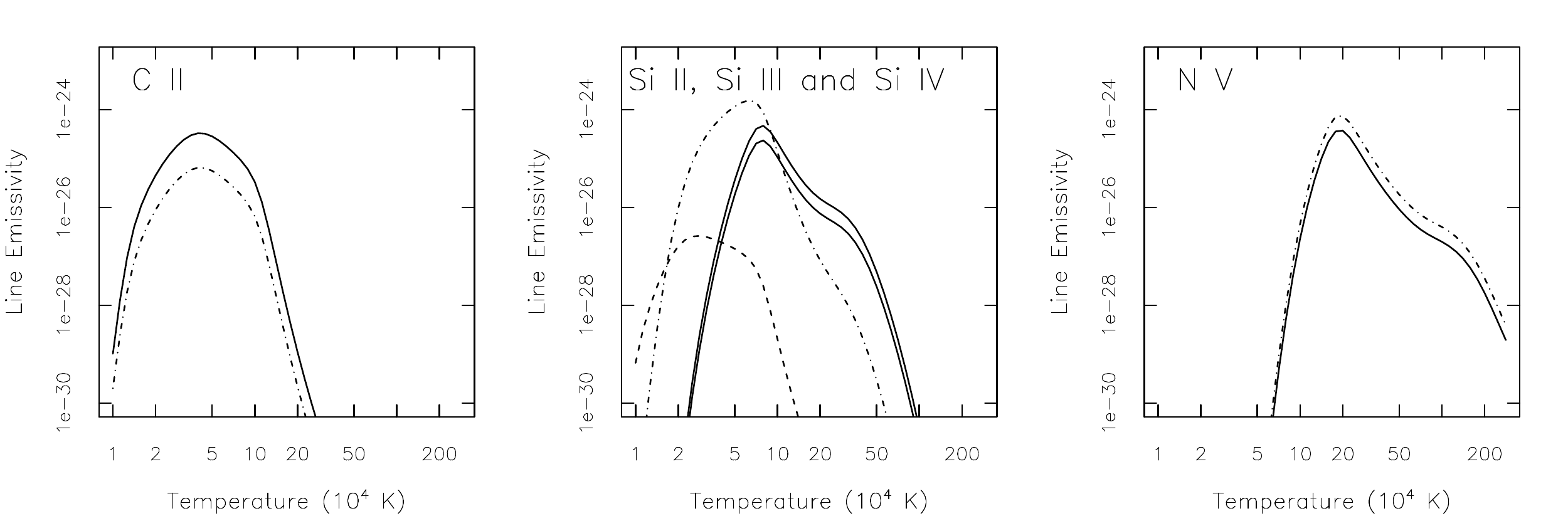}
\caption{\label{emisslines} Emissivity of lines.  in the spectral range $1150\AA-1450\AA$. 
Left: C II doublet. Center:  Si II (dashed line), Si III (dot-dashed line) and Si IV doublet (solid line). 
Right: N V doublet.  
The curves are derived from {\sc CHIANTI} database and {\em ChiantiPy} 
routines. The lines are formed in the range $2-50\times10^4 K$ and thus probe the temperatures
typical of the transition region.}
\end{center}
\end{figure*}

\begin{figure*}
\begin{center}
\includegraphics[width=0.99\textwidth]{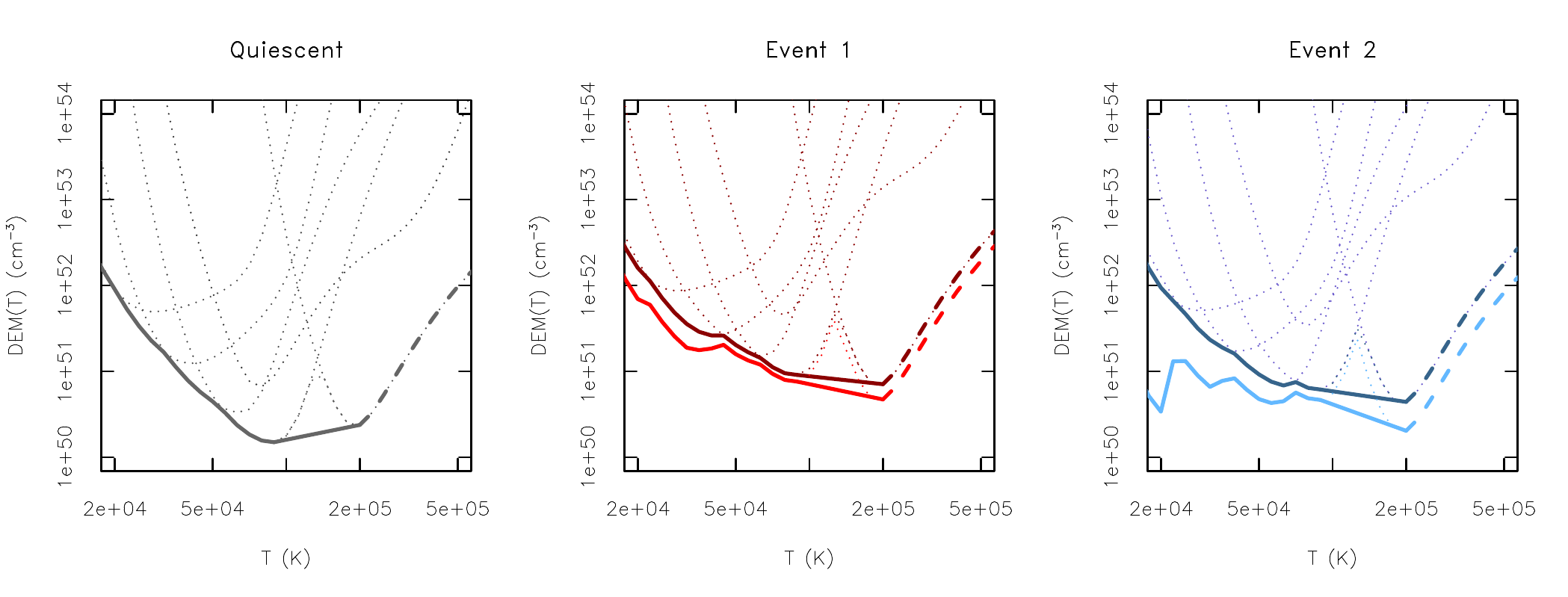}
\caption{\label{dem} Differential emission measure (DEMs) of quiescent spectrum (left panel), first (central
panel) and second brightening (right panel). The dotted curves are the contributions from each line.
DEMs relative to event 1 and 2 are shown as total DEM (dark colored curve), and after subtraction of
the quiescent DEM (light colored curve). For doublets of Si IV, N V and C II we used the sum
of the line fluxes and emissivities. Above 2$\times10^5$ K the curves are constrained only by N~V.}
\end{center}
\end{figure*}

The ion lines in the range of COS spectra have the peaks of formation temperatures 
typical of the transition region, in the 
range $T = 10^4-8\times10^5 \mathrm{K}$ (Fig. \ref{emisslines}). 
The emissivities are calculated by means of 
the {\sc CHIANTI} database and a suite of Python routines {\em ChiantiPy 0.5.2} 
\citep{CHIANTI,Landi2013}. 
We investigated the thermal structure of the emitting 
plasma of \hd\ by means of the differential emission measure (DEMs).
DEMs give a measure of the volume and density of plasma emitting at a certain temperature,
and they are calculated by estimating the lower envelope of the contribution of each line 
(dotted curves in Fig. \ref{dem}).
From the emissivity curves and line fluxes we obtained the DEMs for the quiescent 
part and the two impulsive events (solid thick lines in Fig. \ref{dem}). 
The comparison of the values of the quiescent DEM and of the DEMs of the two brightenings,
and the two slopes around the minimum (occurring at $\sim8\times10^4$ K)
shows that during both events more plasma is contributing to the emitted spectrum in all the
range of temperature. For the first event, the emitting plasma is denser and/or 
fills a bigger volume than in quiescent phase. 
During the second event the slope of the cool side
of the DEM is flatter and steeper than in the quiescent DEM and the DEM
of event 1. This means that relatively hotter plasma is present during event 2, and 
less hot plasma is visible during event 1, remarking again the different nature of the two
brightenings as pointed out before by means of ECDFs. 
The slope of DEMs at high temperatures is poorly constrained, being only due to
the contribution of N V doublet, and thus it is identical in all the three DEMs. 

\subsection{Ly$\alpha$ and other lines}
The main feature in the COS spectra is the H I Ly$\alpha$ line. 
{\ A detailed analysis of  Ly$\alpha$ in spectra of solar type stars 
has been performed by \citet{Wood2005} and, for \hd, by \citet{Bourrier2013}.
Here we attempt to a qualitative description of the  Ly$\alpha$ as observed with COS; 
however, a detailed quantitative analysis is beyond the scope of the paper.

The intrinsic Ly$\alpha$ emission of a star of spectral type of \hd\ is difficult to assess 
because of the core absorption due to interstellar gas.
Without the narrow geocoronal emission and the strong interstellar medium (ISM) absorption, 
the line would appear as a double peaked profile. 
The ISM absorption almost fully depletes the central part of the profile, and, in addition, deuterium 
absorption is present at -80 km/s. 
The contamination from geocoronal emission has a wavelength width comparable with the
COS slit aperture and it is difficult to subtract \citep{Linsky2010}. 
Furthermore, the line shows a periodic change in intensity and profile shape as a function of 
the satellite orbital phase.}
 
We subtracted the airglow emission from Ly$\alpha$ by using a spectrum obtained
during program \# 11999, which has been performed with the same COS setup used in our program.
The airglow spectrum of Ly$\alpha$ has been smoothed with a spline function, corrected for
the reflex satellite movement, and resampled on a grid of 0.04\AA\ centered on the 
nominal line central wavelength (1215.67\AA) in a range of $\pm1.6\AA$ (394.5709 km/s).
The profiles of our spectra were smoothed and resampled in the same way, the airglow spectrum
was scaled to the median of the five bins around the central wavelength and subtracted to our
spectra. The results offer a way to qualitative discuss the ``decontaminated'' Ly$\alpha$ of
\hd\ and compare the profile at the planetary transit and planetary eclipse. 

\begin{figure*}
\includegraphics[width=0.9\textwidth]{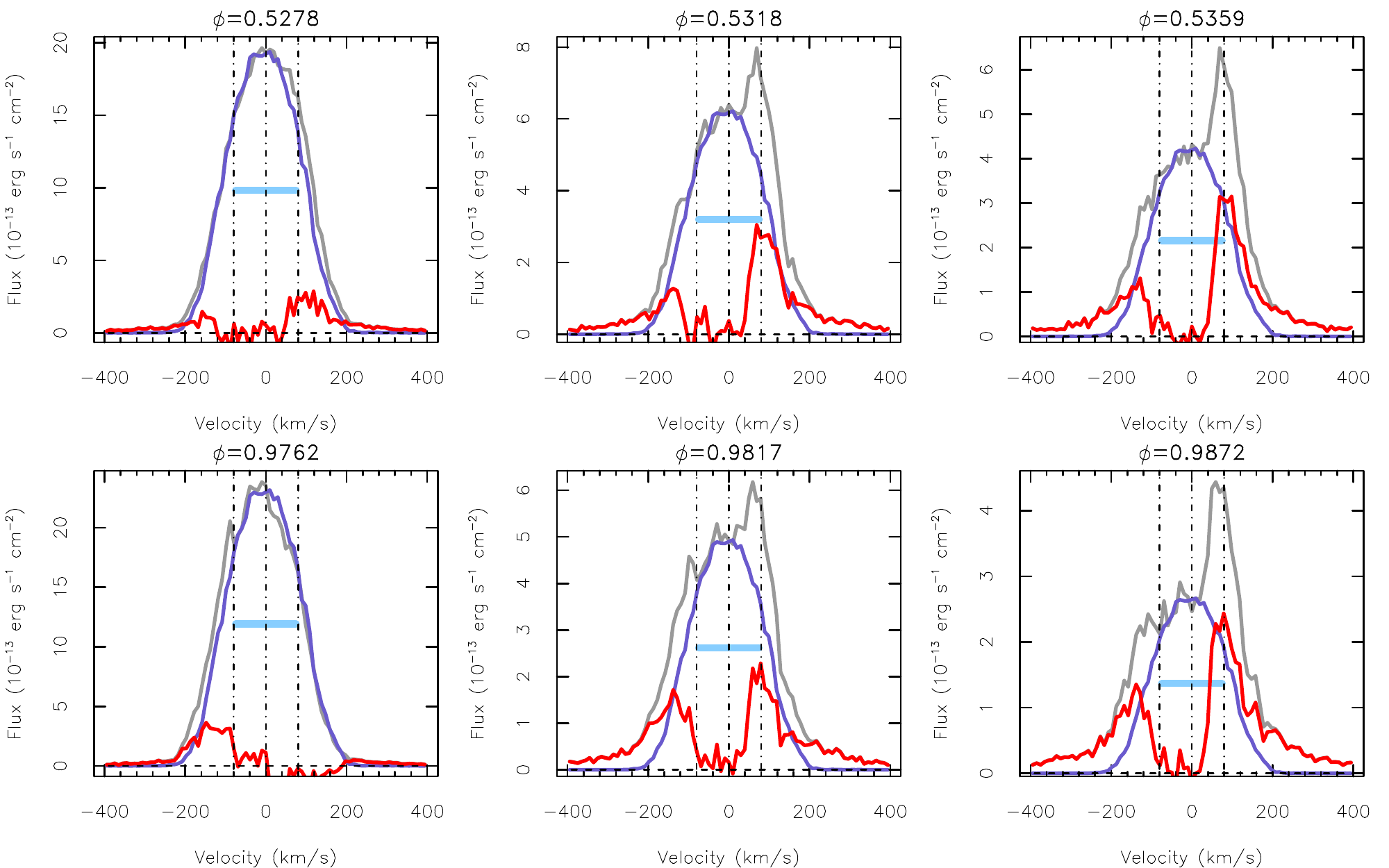}
\caption{\label{lyalpha} Examples of Ly$\alpha$ profiles corrected for the airglow emission
in doppler shift velocities space. Orbital planetary phase is indicated on the top of each 
panel. Gray line is the raw profile, blue is the airglow contamination, red line is the corrected 
profile. Vertical lines mark the rest position, and $\pm80$ km/s. 
The expected width of the uncontaminate and unabsorbed Ly$\alpha$ (150 km/s) 
is shown with {\ the light blue horizontal bar}. }
\end{figure*}

We plot in Fig. \ref{lyalpha} the profiles of  Ly$\alpha$ of \hd\ corrected for airglow
emission, at three epochs, taken after a planetary eclipse and before a transit. 
We chose these phases as representative
of the behavior of the line as a function of the airglow contamination. 
When the airglow emission is at its maximum (left panels in top and bottom rows), 
the correction suffers more of uncertainties, likely due to the saturation of 
the line peak in the detector. 
In the other cases, the correction seems more effective and the 
double peaked profile is recovered. {\ In almost all cases, the red peak is higher than
the blue peak, as observed sometimes in other stars (see Fig. 6 of \citealp{Wood2005})
the reason is likely due to the shape of the interstellar absorption being
more effective in the blue wing of the line while the extra deuterium absorption has
a narrow effect in the blue portion at around -80 km/s \citep{Bourrier2013,Lecavelier2010}.

The average COS spectrum of \hd\ (with an exposure of 12.106 ks) shows other small lines beside the 
Si, C, and N ion lines we have analyzed in details.
}
Lines of Fe XII (1241.8\AA\ and 1349.4\AA ) and Fe XIX (1328.8\AA ) are visible in the 
spectrum of the total exposure. These lines are from hot material composing the corona
of \hd. A few very faint unidentified lines are present at 1274.95\AA , 1289.9\AA , 
1359.2\AA , and 1360.2\AA .

\section{Discussion} \label{discussion}
We have observed \hd\ for five consecutive orbits of HST during which the planet
orbited through the phases $\phi \sim 0.5-0.63$.
We obtained time resolved spectroscopy with COS around Ly$\alpha$ and we
observed two intense line brightenings.
A similar strong variability has been observed previously in the same phase
range in X-rays \citep{Pillitteri2010, Pillitteri2011, Pillitteri2014}.
At these planetary phases the star systematically manifests an extraordinary
rapid variability in the form of FUV brightenings and X-ray flares.
While these events seem typical for an active star, the restricted phase range
in which they manifest is anomalous and strongly points to an activity phased
with the motion of the hot Jupiter around its host star. 
\hd\ behaves markedly active when the planet is emerging from its eclipse.
Of support to this phenomenon is the fact that multiple observations of \hd\ 
obtained at the planetary transits both in FUV and X-rays 
\citep{Poppenhaeger2013, Lecavelier2010} did not
show similar levels of variability (being this of order of $\sim2-3$ sigma level), 
and raising the question of why the star is more
active when the planet is emerging from its eclipse.

\begin{figure*}
\begin{center}
\includegraphics[width=0.9\textwidth]{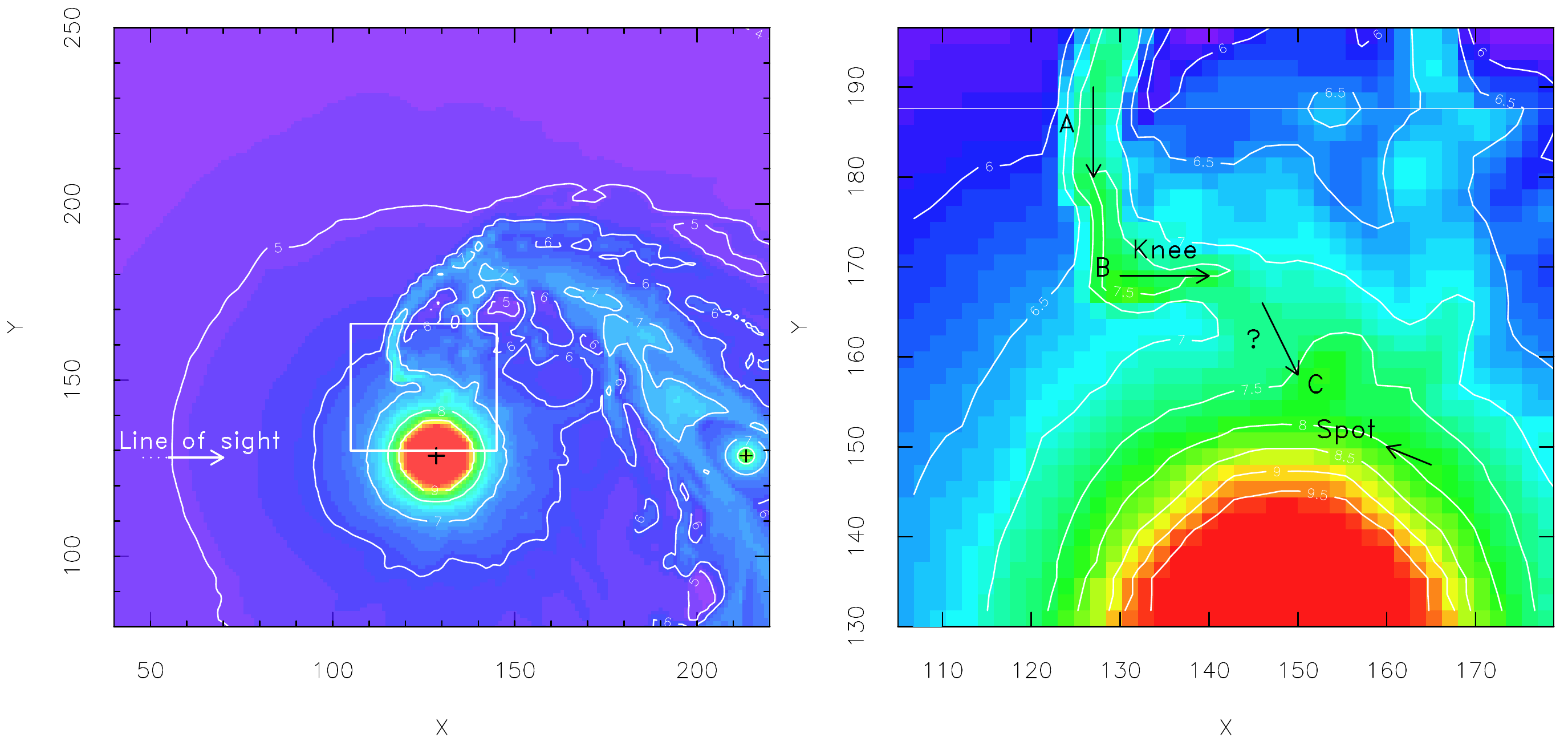}\\
\caption{\label{mhd_stream} 
Particle density contours {\ (units cm$^{-3}$)} of an MHD simulation 
that models star--planet interactions between a hot Jupiter and its host (polar view).
The flow morphology provides a strong candidate scenario to address the FUV line
variability of \hd\ (the observer is viewing the system from the left side).
{\ The star rotates counterclockwise, and the planet orbits the star along the same direction.}
The two ``+'' symbols shown on the left panel indicate the location of the star
(red disk) and the planet (gr een disk).
A close up of the impact region is depicted in the right panel, where the motion
of the accreting plasma is marked with arrows.
Specifically, the shocked plasma is funnelled by the magnetized stellar wind in
an almost radial trajectory close to the star (A), it forms a ``knee'' structure
that consists of hot and dense plasma (B), and then accretes in a spot ahead of
the orbital phase (C).
The precise details of accretion are not investigated by the simulation (zone
marked with {\sc ?}).
The knee (B) of the stream and the active spot upon impact on the surface (C)
are the main sites of production of the enhanced flux observed in FUV and
X-rays that is phased with the orbital motion.}
\end{center}
\end{figure*}

In the following, we propose a scenario where the planet is inducing a
relatively compact hot spot on the stellar surface, which benefits from limb
brightening at specific post-eclipse phases.
In particular, {\ we argue that} such phased FUV brightenings and X-ray flares {\ could be} likely the
signature of star-planet interaction in the form of material evaporating from the planet, 
and accreting onto the star, {\ as suggested by recent 3D magneto-hydrodynamic (MHD) 
simulations} by \citet{Matsakos2015}, ({\ see also \citealp{Cohen2011}}).
{\ The numerical model is setup in PLUTO \citep{Mignone2007,Mignone2012}, 
and includes a magnetized stellar wind, as well as a planetary 
outflow\footnote{\ The simulation does not include photo-evaporation explicitly.
Instead, \citet{Matsakos2015} implemented the planetary outflow profile 
from the detailed simulations, whose results agree with 
\citealt{Murray-Clay2009}). See \citep{Matsakos2015}, for the details of 
the implementation.}, with the stellar
and planetary magnetic fields assumed to be $\sim$$2$\,G and $\sim$$0.1$\,G, respectively.}
The left panel of Fig.~\ref{mhd_stream} shows a snapshot of such a simulation
(polar view) in order to highlight the flow structure.
In {\ this class of systems \citep{Matsakos2015}, submitted, for a classification of
star--planet interactions}), photo-evaporation drives a strong outflow from the surface of
the Hot Jupiter, which becomes supersonic and collides with the stellar wind.
The shocked material is slowed down by the surrounding stellar plasma, and is
dragged inwards by gravity forming a spiral-shaped trajectory.
As a result, the accreting flow impacts onto the stellar surface at a location
that precedes the sub-planetary point.

The right panel of Fig.~\ref{mhd_stream} shows a close-up of the accretion
region.
The effects of the stellar wind and gravity funnel the ionized stream through an
almost radial trajectory towards the star (point A, right panel of
Fig.~\ref{mhd_stream}).
The motion of the infalling material is obstructed by the pressure (thermal plus magnetic) 
of the stellar corona  at a finite height above the stellar
surface (point B).
{\ At that location, the flow forms a hot and dense structure (the ``knee''), with 
number densities up to  $10^8\,$cm$^{-3}$}.\footnote{\ The physical mechanisms
necessary to describe accurately the plasma temperature were computationally
prohibitive given our numerical resources.
Therefore, we do not model the temperature distribution.}
{\ Recent planetary outflow models from \cite{Koskinen2013}  suggest $6-10$ times 
higher mass loss rates than \cite{Murray-Clay2009}, which would increase the
knee density value found in our simulation.
In addition, cooling mechanisms could also contribute to condense the gas.}
Note that the whole spiral-shaped stream corotates with the hot Jupiter, and
hence, both points A and B rotate around the star with an angular velocity
similar to that of the planetary orbit ($P_{\rm pl} \sim 2.2\,{\rm d}$).
However, the material that accumulates in point B experiences the drag of the
slower rotating star ($P_* = 11.9\,{\rm d}$), and as a result, it loses its
angular momentum and lags behind.
In addition, the local pressure distribution pushes the gas to the right,
leading to the formation of the ``knee'' structure highlighted in the right
panel of Fig.~\ref{mhd_stream}.
The material then accretes onto the star, impacting the chromosphere at point C.
{\ A visual inspection suggests that point C precedes the location of the 
planet by about $70-90\deg$.
As the planet orbits around the star, point C is continuously relocated,
effectively corotating with the planet (but the accreted material should 
acquire the rotation of the star after the impact on the surface).
Due to the weaker magnetic fields of Main Sequence stars, the accreting 
plasma is not expected to impact at high latitudes as in protostars.
}

For the discussion that follows, consider the observer (Earth) to be on the left
side of Fig.~\ref{mhd_stream} (the line of sight is depicted with an arrow in
the left panel), thus observing the planet after its eclipse.
At phase $\sim0.52$, the elongated structure of hot material in point B is
approximately aligned with the line of sight.
Since the plasma motion is receding with respect to the observer with a velocity
of $10$--$50\,\rm{km\,s}^{-1}$, its emission will appear redshifted.
Shortly after, the impact region (point C) appears at the stellar limb along the
line of sight.
The hot spot moves towards the observer with the angular velocity of the planet,
corresponding to a local speed of $15$--$20\,\rm{km\,s}^{-1}$ that will
blueshift the emitted radiation.

In this context, we find a plausible match between the accretion features in the
MHD simulations and the FUV observations.
The first brightening can be attributed to the {\em knee} coming into view
approximately when the planet is emerging from the eclipse (point B). 
The observed redshift is in agreement with the simulation where the flow is
receding along the line of sight.
Because of the shock which is likely to form there, the gas in the knee is
expected hotter than the average temperature of the transition region, but still
cooler than the gas impacting on the surface of the star.

The second brightening can be attributed to the spot emerging later at the limb,
which is due to the impact of the accreting gas onto the chromosphere.
It is plausible to imagine that the accreting spot is forced to move with the
co-moving speed of the planet at the stellar surface ($\sim18.3$ km/s), since the
stream inherits the angular speed of the orbit.
As a result, a blueshift of the lines is observed because the spot is moving
toward the observer, and at a rate that is faster than the rotational velocity
of the stellar surface (3.4 km/s). 

Note that secondary flows may detach from the main body of the stream and fall
onto the surface with minor impacts.
Nevertheless, the spot of the bulk accretion remains confined in a restricted
range of planetary orbital phases, and remains within the range of phases in
which we observed variability in \xmm\ and in the present HST observations.

The accreting stream is moving in an elongated magnetized region (marked by
points A and B of Fig.~\ref{mhd_stream}) that can host magnetic reconnection
(and thus flaring activity).
The relevant size is markedly different from the typical sizes of coronal loops 
of Main Sequence stars like the Sun. 
An estimate of the size of such long structures has been inferred from the 
analysis of the decay light curve of a flare observed in X-rays 
\citep{Pillitteri2014}, and it is of a few stellar radii. 
From the simulations, we estimate that the stream forms an arc before the knee of
the size of the $1.5-2$ stellar radii, and the knee is about $1.5$ stellar radii
above the stellar surface. The mass flux through the knee is estimated of order of
a few units of {\ $10^{-8}$ gr s$^{-1}$ cm$^{-2}$}.
From this mass flux, assuming an accreting tube with section of $0.1\times R_*$,
we estimate a mass accretion from the planet of {\ $\sim5\times10^{18}$ gr yr$^{-1}$}, 
or up to about {\ 1\%} of a Jupiter mass in $\sim4$ Gyr, if the accretion were steady during this period of time.

We warn that the numerical model does not have adequate resolution to model in
detail the path of the plasma flowing from point B to C (limited by the
demanding nature of 3D MHD simulations).
The precise location of the spot (point C) may vary slightly due to the
time-dependence of the plasma dynamics.
It is plausible that the accretion is not happening in form of a continuous
flow, but instead, the plasma accumulates until the density of the gas excedes
the confining pressure of the magnetic field, and mimicks the fall of
dense fragments observed just after the solar eruption of June 7 2011
\citep{Reale2013, Reale2014}.
The limitations of the numerical simulation prevented us also from following in
detail the impact onto the stellar surface.
The surface is treated as a boundary which does not physically describe the
chromosphere, but rather the base of the stellar wind.

Such MHD simulations of systems harboring hot Jupiters differ from previous
simulations (e.g., \citealp{Cohen2011}) in the treatment of the stellar and
planetary winds.
They also differ from analytical models of purely magnetic star-planet
interaction \citep{Lanza2009, Lanza2010, Lanza2011}.
However, all these modeling attempts concur to qualitatively describe a
preferred planetary phase range in which stellar activity is enhanced by
magnetic reconnection, or, by accreting material, as in the model presented
above from \citet{Matsakos2015}.

{\ Stars accrete material from their circumstellar disks during the pre- Main Sequence phase,
and, depending on the rate of accretion, they are identified as Classical T-Tauri (
CTTSs, strong accretors) and Weak-line T-Tauri (WTTSs, weak accretors). 
Given that we make the hypothesis that \hd\ is accreting material from its hot 
Jupiter, a comparison between the spectrum of \hd\ and CTTSs/WTTSs stars is in order.  
\citet{Ardila2013} present an extended and thorough analysis of FUV spectra of a
sample of CTTSs and WTTSs.
They recognize a broad and a narrow components in the C IV, Si IV and N V lines,
with the relative weight of the narrow to broad component varying between 20\%
and 80\% as a function of the accretion rate.
The lines of WTTSs are shaped mainly by a low redshifted, narrow component. 
C IV lines are redshifted in both CTTSs and WTTSs, with the redshift being of
order of 10 km/s in WTTSs. FUV lines of \hd\ differ from those of CTTSs and WTTSs.
In \hd\ we observe narrow lines like in WTTSs and 
a systematic redshift of C II lines that is usually associated with emission
from the gas flowing down in the coronal loops.

{\ We find similarities between the spectral features of \hd\ and 
the FUV spectrum of the solar eruption of June 7 2011.}
The solar event has been analyzed in detail by \citet{Reale2014}. 
This unusual eruption showed the break of a dense filament that ejected dense 
fragments eventually falling down  on the solar surface. 
The dynamics of this eruption is the best template for accreting events
in pre Main Sequence stars \citep{Reale2013} and, in our case, for \hd.
The brightenings of C IV doublet at 1550\AA\ were modeled by \citep{Reale2014} 
by means of a cloud of fragments hitting the top of the chromosphere. 
Among the results, \citet{Reale2014} explain the broad component of FUV lines
in accreting stars as due to the accretion taking place simultaneously
at different locations on the stellar surface. 
They also do not find a continuum increase in the reconstructed spectra of the 
impacting fragments. 
In solar flares due to magnetic reconnection, a rise of the continuum flux is observed 
as due to the heating process and the rise of the free-bound and free-free contributions \citep{Brekke1996}. 
{\ The lack of continuum flux increase in the brightenings we observed in \hd\ is analog to 
the solar behavior during the impacts of the dense fragments after their eruption. }
By means of ECDFs we have established in Sect. \ref{secondevent} that the second event has 
an impulsive nature. The lack of continuum increase during this brightening
is analog to the lack of continuum increase in the solar event described above. 
This allows us to consider the second brightening in \hd\ as an accretion event of
planetary material similar to the solar event rather than a flare due to magnetic 
interaction and/or reconnection. 
On the other hand, the first event is rather due to the emergence of a hotter, 
elongated region (knee) briefly seen through a favorable line of sight. 
}

On the wider context of systems with hot Jupiters, 
the conditions for establishing a steady accreting stream 
strongly depend on {\ the stellar wind}, the intensity {\ and geometry} of the stellar 
magnetic field, the stellar activity (related to the magnetic activity), and the separation
of star and planet (which controls also the rate of evaporation of the 
planetary atmosphere). In this respect, \hd\ demonstrates to be one of the most 
favorable systems where this phenomenon can be detected. 
The presence of a multi-year activity cycle {\ (see, e.g., \citealp{Fares2013}),
associated with a variability of reconnection events \citep[e.g.][]{Llama2013}},
implies alternating periods of active accretion and periods of weak or absent accretion. As a consequence,
an activity cycle would explain lack of systematic phased variability when
observing stars after a few years, and the on/off behavior observed 
in the case of HD~179949 \citep{Shkolnik08}. 

{\ We find unlikely that the two events are related to hot material emitting close to the planet.
Such material should be ejected from the planet and inherit a velocity of at least 150 km/s which
is the orbital velocity of the planet. Similar velocity should be involved in magnetic reconnection
events close to the planet. 
At phases 0.52 and 0.59 the components of such velocity along the line of sight are $\ge-18$ km/s and 
$\ge -80$ km/s, respectively. In both cases we should see a blue shift when observing after
the planetary eclipse. At phase 0.52 we observe a redshift of about +20 km/s, 
thus a difference of $\ge40$ km/s with respect
to the expected Doppler shift of emission from gas close to the planet. 
For the second event at  phase 0.59, the expected blue shift of -80 km/s 
from material close to the planet. Such blue shift is not detected in any of the lines.

SPI events should be synced with the synodic period of the star+planet system which is 
about 2.7 days.
This value is quite close to the orbital period of the planet, thus disentangling this effect
from other effects purely related to the planetary motion is difficult. 
In addition it would require a continuous monitoring of the star for at least 3 synodic 
periods, which is a demanding observational effort not realized to date.

}
\section{Conclusions} \label{conclusions}
In this paper we have presented a time resolved spectroscopic analysis of the
FUV spectrum of \hd\ acquired with COS on board HST, around the Ly$\alpha$ line 
in the range 1150-1450\AA. 
We have observed the phases after the planetary eclipse, for a duration of
five HST orbits. The spectral range comprises several lines of ionized Si, C, N and O,
which form in the range of temperature of $20,000-200,000$ K, typical of the 
transition region.

Two rapid line flux brightenings have been observed in Si, C, and N lines with
a duration of $200-400$ s. No similar brightenings have been observed in archival 
observations of \hd\ taken with the same instrument at a planetary transit. 
The first event happens when the planet is emerging after 
the eclipse (phase $\sim0.52$), the second event later at phase $\sim0.59$.
From the line ratios of Si ions, and with a reconstruction of the  
differential emission measure, we followed the evolution of the plasma temperature
and the rise of temperature during both events. 
The time evolution of the line fluxes is different and hint to a different nature of the two events: 
the evolution of the first event has been gradual within 400 s. The second event has been 
a rapid impulse observed first in the hot lines of N~V, and then in the cooler lines of Si IV and C II.
The second event has been also hotter than the first one. For the second event,
we find a strong similarity with the impacts of dense fragments during the 
solar eruption of June 7 2011. In particular, both events are impulsive
and lack of continuum increase, at odds with solar and stellar flares due to
magnetic reconnection.

With the help of MHD simulations, {\ we argue that the spectroscopic sequence is the
signature of gas that evaporates from the planet and accretes onto the star.} 
In particular, the MHD simulations show that material {\ can effectively escape} 
the planetary surface, becomes supersonic, and collides with the stellar wind.
The interaction leads to the formation of an elongated spiral-shaped stream that
precedes the planetary motion, and a cometary-type tail that trails the orbit.
Near the star, the stream is funneled by the surrounding magnetized plasma, and
infalls almost radially.
The flow is then obstructed by the coronal pressure at a finite height above the
stellar surface, possibly forming a shock.
The dynamical evolution then leads to the formation of a ``knee'' structure of
dense and hot plasma, that is subsequently channeled inwards and impacts onto
the stellar surface.
As a consequence, the knee region and the surface accretion spot are hotter and
denser than the average transition region of the star.
Since the entire accreting stream is co-rotating with the planet, the phased
variability occurs when -- under favorable alignment with the line of sight --
the knee and the accretion spot emerge at the stellar limb.
The phase angle between the accretion spot and the sub-planetary point matches
quite well the observed phase lag in X-ray and FUV activity.

This scenario is appealing because it can explain both X-ray and FUV observation after the 
planetary eclipse. These observations provide the spectroscopic signatures of the accreting planetary 
gas. However, further monitoring at the post planetary eclipse phases {\ and on longer time baseline} 
are needed to confirm the systematic occurrence of such spectroscopic signatures and the enhanced 
activity of \hd\ at these particular phases. 

\acknowledgments
IP is grateful to H. M. G\"unther for the  help and the discussion of the results. 
IP acknowledge financial support from the European Union Seventh Framework Programme (FP7/2007-2013) 
under grant agreement n$^{\rm o}$ 267251 “Astronomy Fellowships in Italy” (AstroFIt).
TM was supported in part by NASA ATP grant NNX13AH56G.
The simulation was carried out with resources provided by the University of
Chicago Research Computing Center.
Based on observations made with the NASA/ESA Hubble Space Telescope, obtained at the Space 
Telescope Science Institute, which is operated by the Association of Universities for Research 
in Astronomy, Inc., under NASA contract NAS 5-26555. 
These observations are associated with program \#12984. Support for program \#12984 was provided 
by NASA through a grant from the Space Telescope Science Institute, which is operated by the 
Association of Universities for Research in Astronomy, Inc., under NASA contract NAS 5-26555.
 
\appendix

\begin{figure}
\includegraphics[width=0.7\columnwidth]{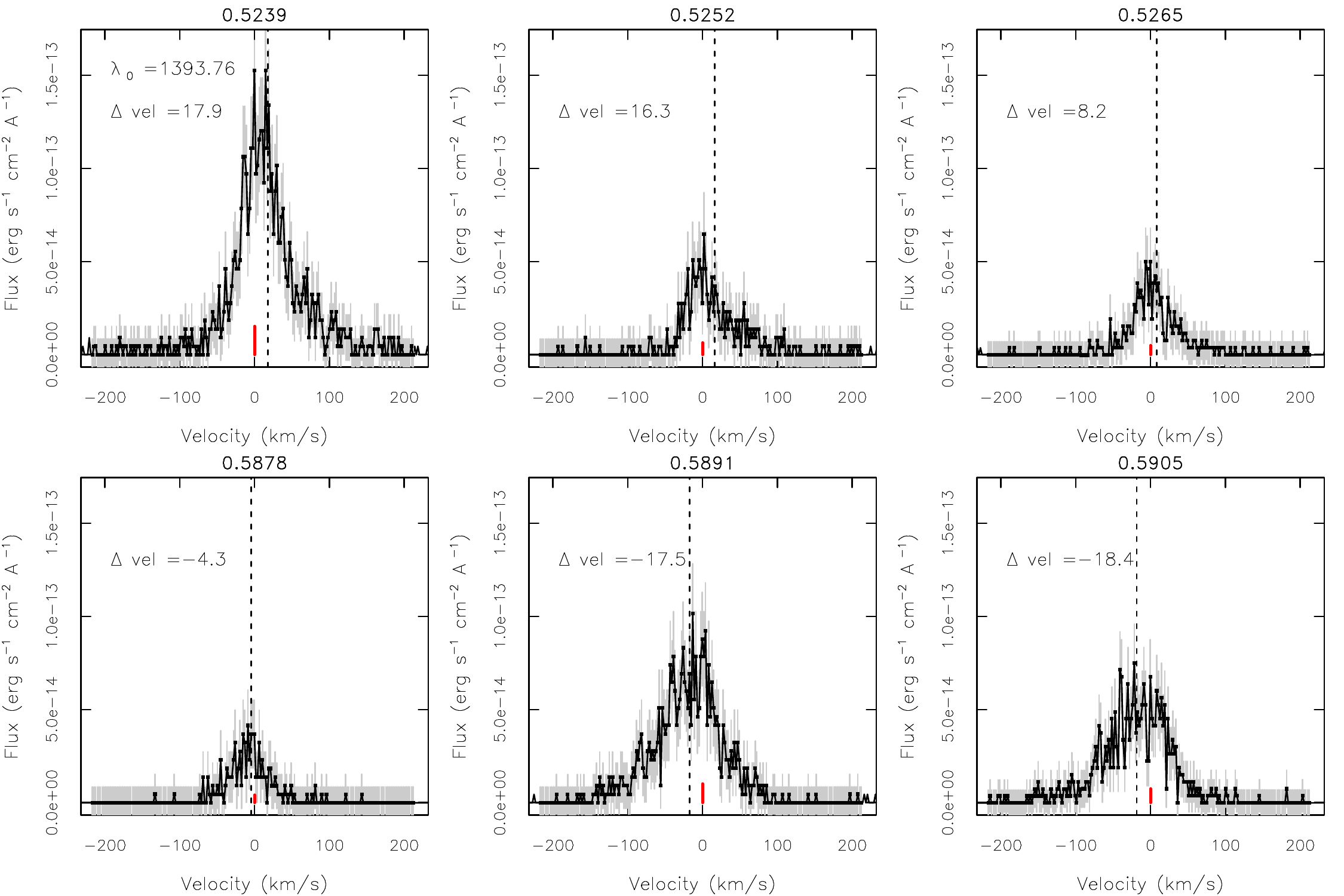}
\caption{\label{si_iv_profiles} Line profiles of Si IV at 1393.7\AA. We
show here the spectra from the three sub-exposures  of $\sim200$ s 5a, 5b, 5c, 14a, 14b, 14c. 
The values of the planetary phases are indicated on the top axis of each panel. 
The vertical line marks the centroid of the line,
the red segment is the rest line wavelength (value shown in the top left panel). The differences
between centroid and rest wavelength in velocity space (km/s) are indicated in each panel
(corrected for stellar radial velocity, RV = 2.6 km/s).}
\end{figure}

\begin{figure}
\includegraphics[width=0.7\columnwidth]{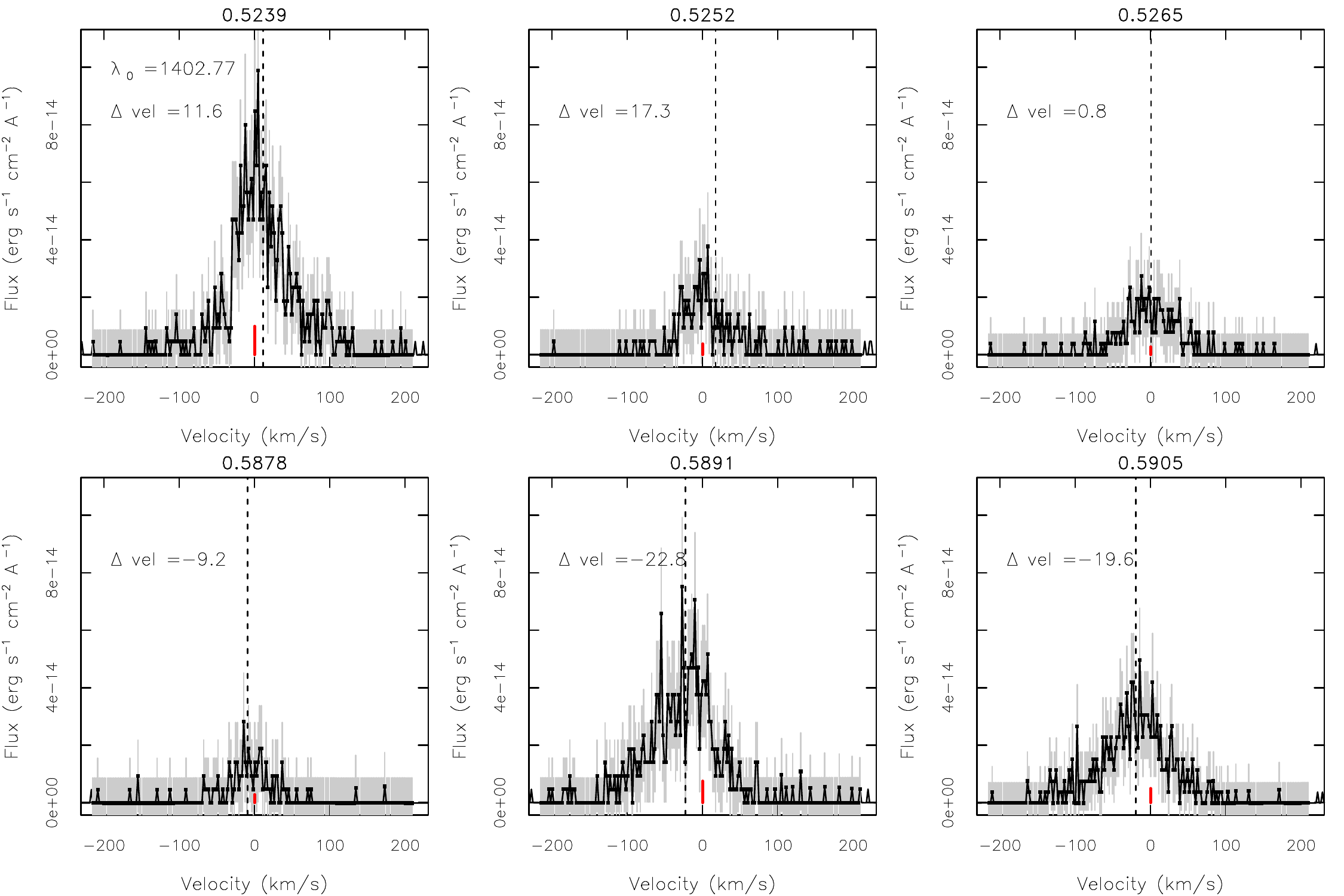}
\caption{\label{si_iv2_profiles} Same as in Fig. \ref{si_iv_profiles} for the Si IV line
at 1402.8 \AA.}
\end{figure}

\begin{figure}
\includegraphics[width=0.7\columnwidth]{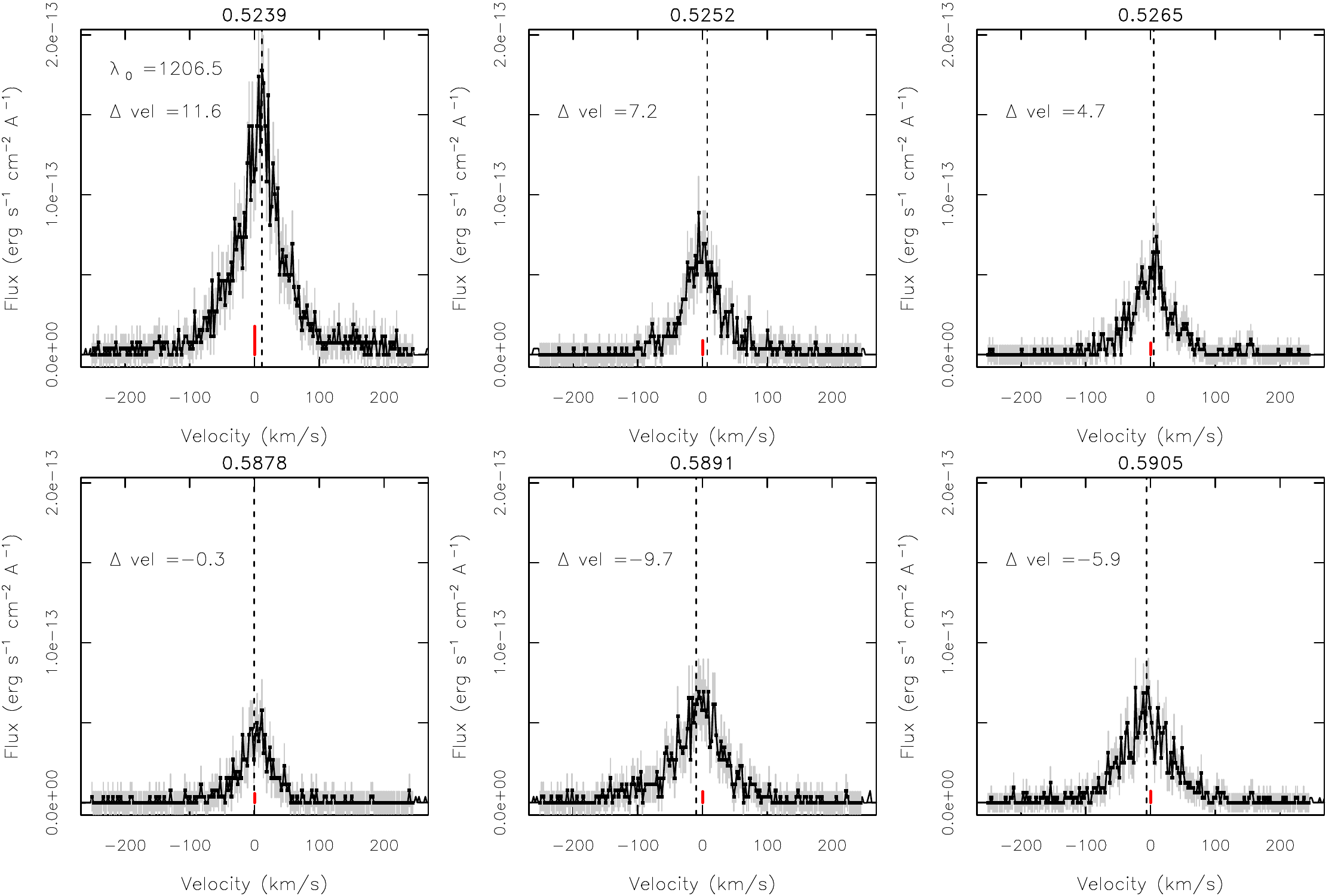}
\caption{\label{si_ii_profiles} Same as in Fig. \ref{si_iv_profiles} for the Si III line
at 1206.5\AA.}
\end{figure}

\begin{figure}
\includegraphics[width=0.7\columnwidth]{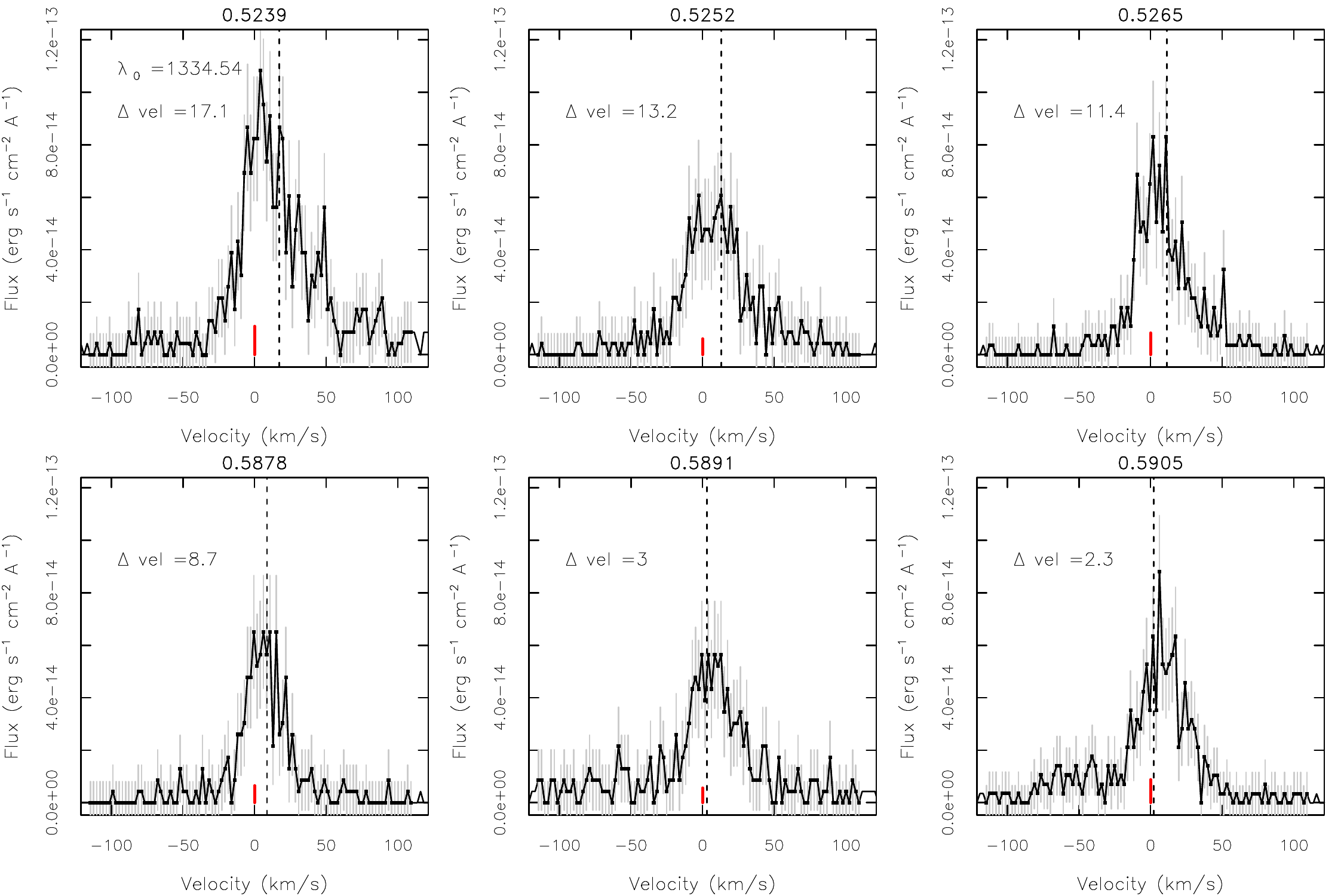}
\caption{\label{c_ii_profiles} Same as in Fig. \ref{si_iv_profiles} for the C II line
at 1334.5\AA.}
\end{figure}

\begin{figure}
\includegraphics[width=0.7\columnwidth]{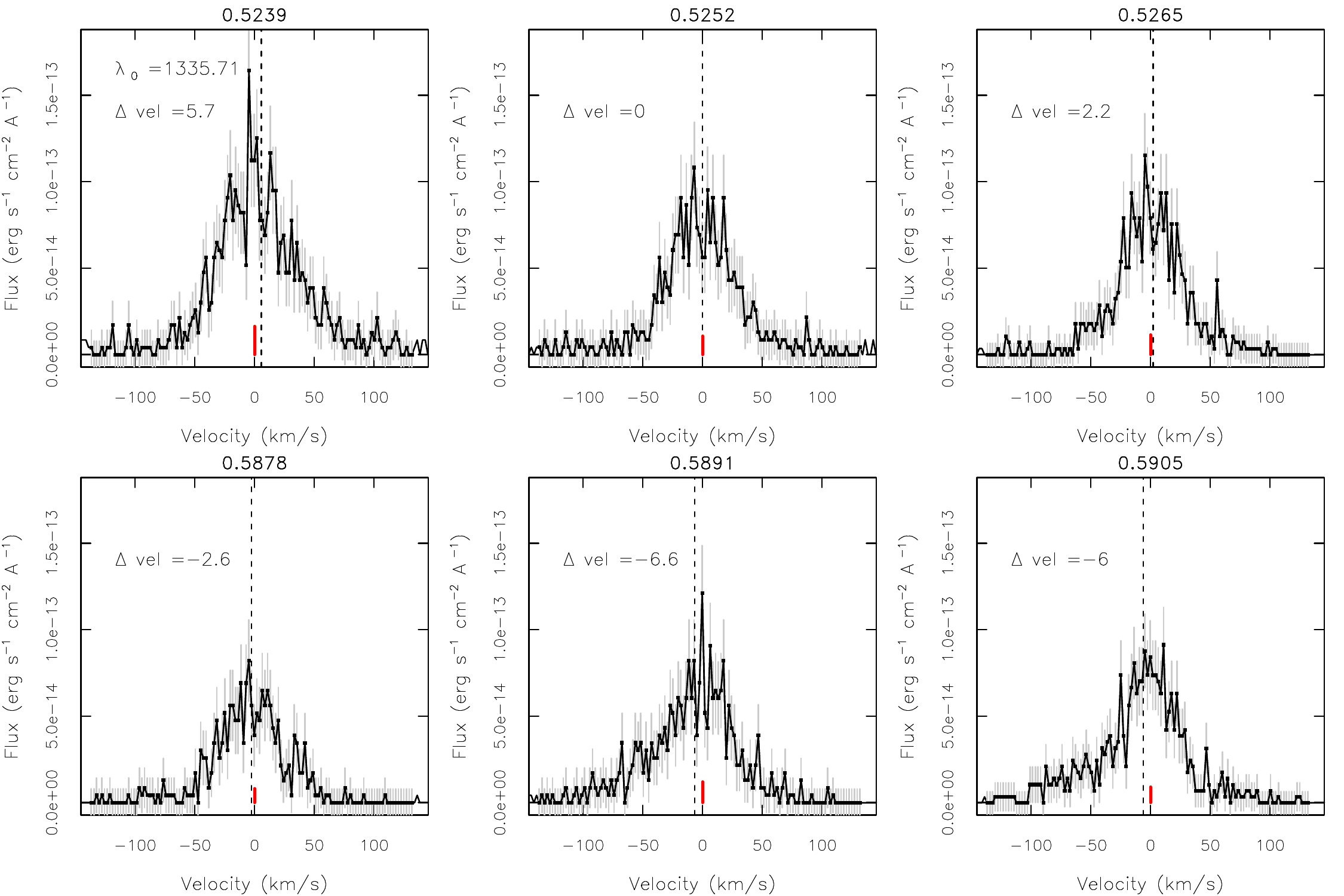}
\caption{\label{c_ii2_profiles} Same as in Fig. \ref{si_iv_profiles} for the C II line
at 1335.7\AA.}

\end{figure}

\begin{figure}
\includegraphics[width=0.7\columnwidth]{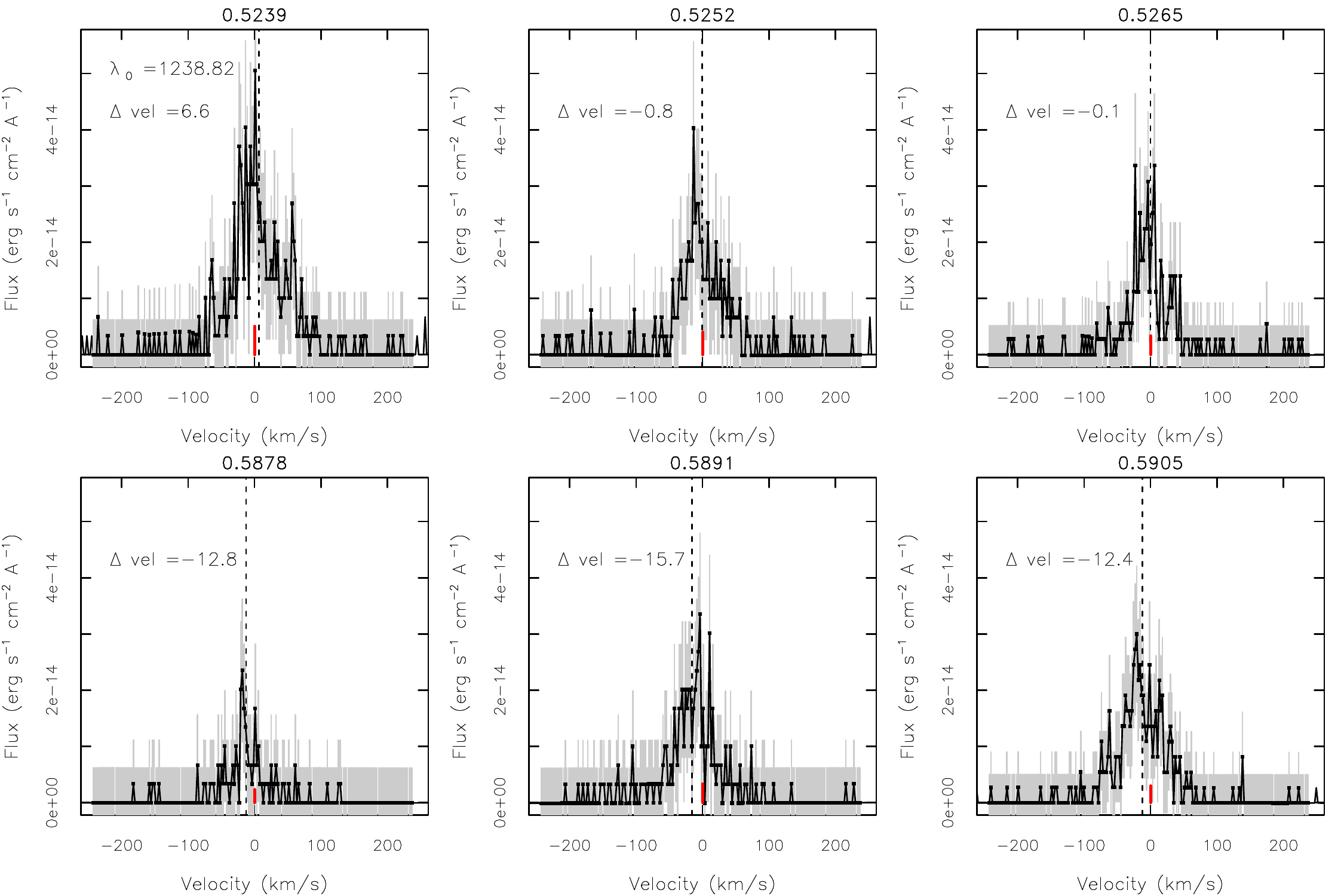}
\caption{\label{n_v_profiles} Same as in Fig. \ref{si_iv_profiles} for the N V line
at 1238.8\AA.}
\end{figure}

\begin{figure}
\includegraphics[width=0.7\columnwidth]{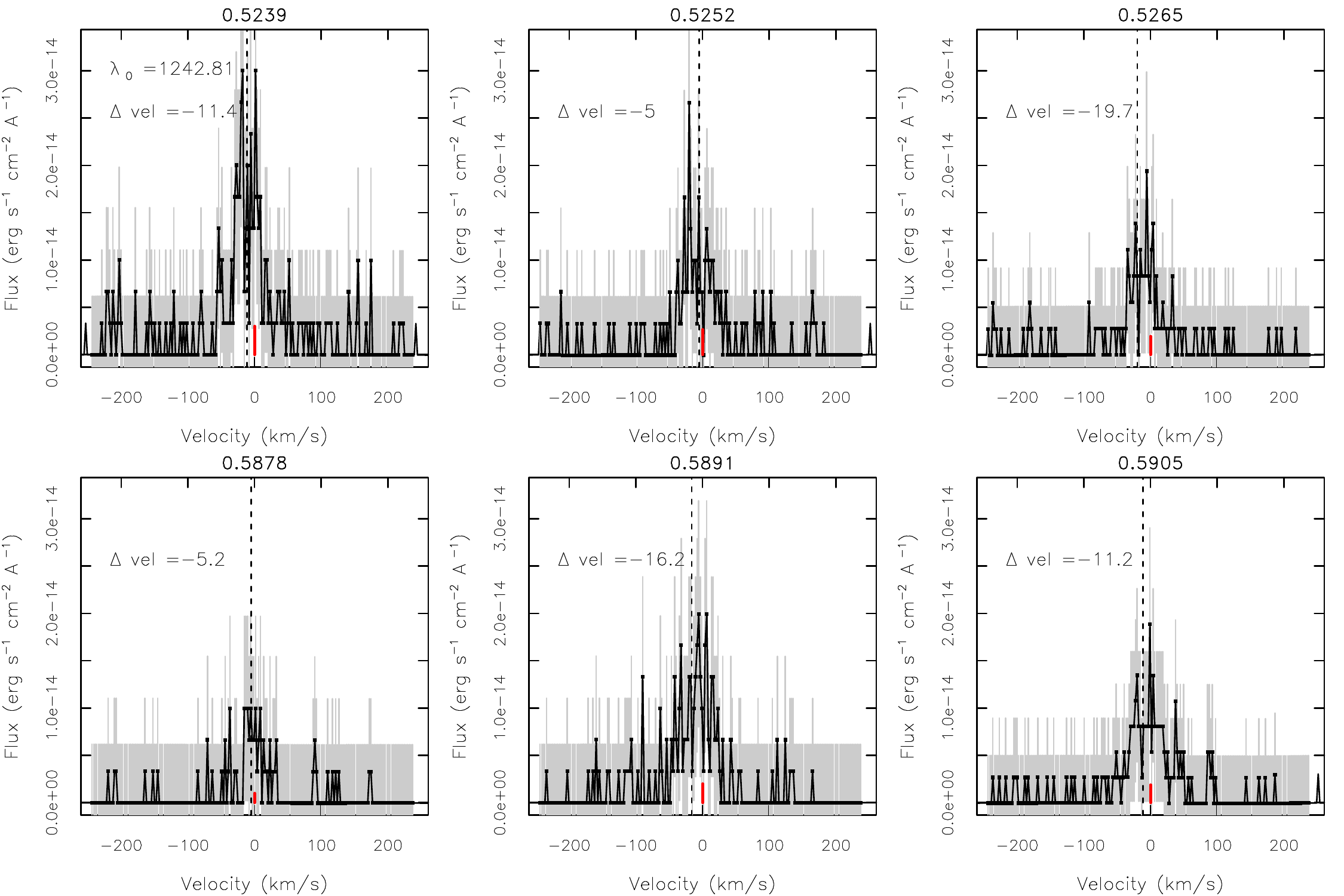}
\caption{\label{n_v_profiles2} Same as in Fig. \ref{si_iv_profiles} for the N V line
at 1242.8\AA.}
\end{figure}

\begin{table}
\centering
\caption{\label{lineflu1} 
Centroids, FWHMs and fluxes of Si IV and N V lines for each exposure.}
\resizebox{0.99\textwidth}{!}{
\begin{tabular}{l|rrrr|rrrr|rrrr}
  \hline
\hline
Exposure & Wavelength & FWHM & Flux & Error & Wavelength & FWHM & Flux & Error & Wavelength & FWHM & Flux & Error \\ 
         & \AA\  & \AA\ & \multicolumn{2}{c|}{$10^{-15}$ erg cm$^{-2}$ s$^{-1}$} &  \AA\  & \AA\ & \multicolumn{2}{c|}{$10^{-15}$ erg cm$^{-2}$ s$^{-1}$} &  \AA\  & \AA\ & \multicolumn{2}{c}{$10^{-15}$ erg cm$^{-2}$ s$^{-1}$} \\
&  \multicolumn{4}{c|}{Si IV \@ 1393.7 \AA\ } & \multicolumn{4}{c|}{Si IV \@ 1402.7 \AA\ } & \multicolumn{4}{c}{Si II \@ 1264.7 \AA\ } \\
  \hline
  1 & 1393.74 & 0.10 & 7.13 & 0.76 & 1402.77 & 0.19 & 3.36 & 0.68 & 1264.70 & 0.18 & 2.10 & 0.46 \\ 
  2 & 1393.74 & 0.25 & 6.78 & 0.74 & 1402.75 & 0.19 & 3.41 & 0.67 & 1264.78 & 0.25 & 2.33 & 0.46 \\ 
  3 & 1393.72 & 0.19 & 6.42 & 0.73 & 1402.73 & 0.18 & 3.26 & 0.66 & 1264.84 & 0.40 & 1.89 & 0.44 \\ 
  4 & 1393.71 & 0.21 & 6.43 & 0.73 & 1402.72 & 0.13 & 3.63 & 0.67 & 1264.76 & 0.35 & 2.18 & 0.45 \\ 
  5 & 1393.83 & 0.23 & 23.16 & 0.78 & 1402.82 & 0.29 & 13.05 & 0.66 & 1264.77 & 0.33 & 3.01 & 0.35 \\ 
  5a & 1393.84 & 0.26 & 45.62 & 2.14 & 1402.83 & 0.25 & 25.11 & 1.84 & 1264.75 & 0.17 & 3.69 & 0.97 \\ 
  5b & 1393.83 & 0.17 & 14.62 & 1.61 & 1402.85 & 0.19 & 8.10 & 1.48 & 1264.76 & 0.44 & 2.84 & 0.94 \\ 
  5c & 1393.80 & 0.21 & 11.56 & 1.33 & 1402.78 & 0.32 & 7.13 & 1.25 & 1264.79 & 0.52 & 2.59 & 0.79 \\ 
  6 & 1393.79 & 0.17 & 9.16 & 0.58 & 1402.77 & 0.29 & 4.79 & 0.51 & 1264.77 & 0.42 & 2.14 & 0.32 \\ 
  7 & 1393.75 & 0.16 & 7.21 & 0.55 & 1402.78 & 0.08 & 3.78 & 0.48 & 1264.80 & 0.32 & 2.10 & 0.32 \\ 
  8 & 1393.75 & 0.20 & 6.85 & 0.54 & 1402.77 & 0.14 & 3.83 & 0.49 & 1264.81 & 0.42 & 2.20 & 0.32 \\ 
  9 & 1393.73 & 0.17 & 6.40 & 0.54 & 1402.76 & 0.28 & 3.37 & 0.48 & 1264.75 & 0.39 & 2.02 & 0.32 \\ 
  10 & 1393.73 & 0.16 & 6.82 & 0.54 & 1402.74 & 0.06 & 3.07 & 0.48 & 1264.79 & 0.39 & 1.84 & 0.32 \\ 
  11 & 1393.73 & 0.16 & 7.86 & 0.56 & 1402.75 & 0.15 & 4.06 & 0.50 & 1264.78 & 0.10 & 2.10 & 0.32 \\ 
  12 & 1393.72 & 0.23 & 7.10 & 0.55 & 1402.74 & 0.30 & 3.63 & 0.49 & 1264.80 & 0.20 & 1.93 & 0.32 \\ 
  13 & 1393.74 & 0.22 & 6.36 & 0.53 & 1402.74 & 0.24 & 2.86 & 0.47 & 1264.76 & 0.37 & 2.23 & 0.33 \\ 
  14 & 1393.68 & 0.31 & 22.26 & 0.77 & 1402.68 & 0.30 & 13.45 & 0.66 & 1264.75 & 0.29 & 2.37 & 0.33 \\ 
  14a & 1393.74 & 0.26 & 8.16 & 1.46 & 1402.73 & 0.11 & 4.36 & 1.39 & 1264.79 & 0.30 & 2.06 & 0.91 \\ 
  14b & 1393.68 & 0.26 & 33.02 & 1.95 & 1402.67 & 0.29 & 20.76 & 1.76 & 1264.68 & 0.50 & 2.17 & 0.92 \\ 
  14c & 1393.67 & 0.43 & 24.96 & 1.55 & 1402.68 & 0.59 & 14.91 & 1.40 & 1264.76 & 0.13 & 2.78 & 0.78 \\ 
  15 & 1393.75 & 0.14 & 8.13 & 0.57 & 1402.73 & 0.18 & 4.39 & 0.50 & 1264.77 & 0.35 & 2.04 & 0.32 \\ 
  16 & 1393.73 & 0.17 & 6.63 & 0.54 & 1402.75 & 0.16 & 3.73 & 0.49 & 1264.83 & 0.43 & 2.26 & 0.33 \\ 
  17 & 1393.76 & 0.12 & 6.58 & 0.54 & 1402.76 & 0.16 & 3.63 & 0.49 & 1264.75 & 0.45 & 2.33 & 0.33 \\ 
  18 & 1393.75 & 0.20 & 7.43 & 0.56 & 1402.73 & 0.20 & 3.73 & 0.49 & 1264.77 & 0.12 & 2.25 & 0.33 \\ 
  19 & 1393.75 & 0.15 & 6.18 & 0.53 & 1402.76 & 0.05 & 3.21 & 0.48 & 1264.77 & 0.09 & 2.00 & 0.32 \\ 
  20 & 1393.75 & 0.19 & 6.53 & 0.54 & 1402.73 & 0.25 & 3.21 & 0.48 & 1264.81 & 0.11 & 2.25 & 0.33 \\ 
  \hline
  \end{tabular}
}
\end{table}

\begin{table}
\centering
\caption{\label{lineflu2} 
Line measurements as in Table \ref{lineflu1} for lines of Si III and C II doublet.} 
\resizebox{0.99\textwidth}{!}{
\begin{tabular}{l|rrrr|rrrr|rrrr}
  \hline
\hline
Exposure & Wavelength & FWHM & Flux & Error & Wavelength & FWHM & Flux & Error & Wavelength & FWHM & Flux & Error \\ 
         & \AA\  & \AA\ & \multicolumn{2}{c|}{$10^{-15}$ erg cm$^{-2}$ s$^{-1}$} &  \AA\  & \AA\ & \multicolumn{2}{c|}{$10^{-15}$ erg cm$^{-2}$ s$^{-1}$} &  \AA\  & \AA\ & \multicolumn{2}{c}{$10^{-15}$ erg cm$^{-2}$ s$^{-1}$} \\
 & \multicolumn{4}{c|}{Si III \@ 1206.5 \AA\ } & \multicolumn{4}{c|}{C II \@ 1334.6 \AA\ } & \multicolumn{4}{c}{C II \@ 1335.7 \AA\ } \\
  \hline
  1 & 1206.51 & 0.18 & 11.76 & 0.73 & 1334.57 & 0.10 & 9.45 & 0.65 & 1335.70 & 0.17 & 17.30 & 0.81 \\ 
  2 & 1206.54 & 0.13 & 11.31 & 0.71 & 1334.58 & 0.13 & 8.83 & 0.63 & 1335.71 & 0.21 & 18.00 & 0.81 \\ 
  3 & 1206.54 & 0.14 & 10.63 & 0.70 & 1334.57 & 0.09 & 8.59 & 0.64 & 1335.72 & 0.17 & 17.64 & 0.80 \\ 
  4 & 1206.53 & 0.19 & 12.35 & 0.73 & 1334.57 & 0.14 & 9.17 & 0.63 & 1335.70 & 0.17 & 17.52 & 0.80 \\ 
  5 & 1206.54 & 0.24 & 29.16 & 0.76 & 1334.60 & 0.15 & 15.87 & 0.60 & 1335.72 & 0.21 & 28.72 & 0.77 \\ 
  5a & 1206.55 & 0.19 & 53.50 & 1.99 & 1334.61 & 0.25 & 21.74 & 1.44 & 1335.74 & 0.18 & 36.47 & 1.75 \\ 
  5b & 1206.53 & 0.18 & 20.12 & 1.49 & 1334.59 & 0.16 & 13.61 & 1.26 & 1335.71 & 0.25 & 25.86 & 1.56 \\ 
  5c & 1206.52 & 0.17 & 16.43 & 1.23 & 1334.59 & 0.14 & 12.86 & 1.08 & 1335.72 & 0.20 & 24.66 & 1.35 \\ 
  6 & 1206.54 & 0.25 & 14.97 & 0.59 & 1334.58 & 0.11 & 11.77 & 0.53 & 1335.71 & 0.19 & 21.99 & 0.68 \\ 
  7 & 1206.53 & 0.13 & 11.75 & 0.54 & 1334.58 & 0.15 & 10.39 & 0.52 & 1335.71 & 0.19 & 20.72 & 0.66 \\ 
  8 & 1206.52 & 0.25 & 11.74 & 0.55 & 1334.57 & 0.10 & 10.72 & 0.51 & 1335.69 & 0.20 & 20.50 & 0.66 \\ 
  9 & 1206.50 & 0.15 & 11.45 & 0.54 & 1334.57 & 0.10 & 8.98 & 0.48 & 1335.69 & 0.24 & 17.90 & 0.63 \\ 
  10 & 1206.54 & 0.14 & 11.14 & 0.54 & 1334.57 & 0.13 & 8.81 & 0.48 & 1335.71 & 0.23 & 18.42 & 0.64 \\ 
  11 & 1206.52 & 0.15 & 13.55 & 0.57 & 1334.56 & 0.15 & 9.81 & 0.52 & 1335.71 & 0.18 & 19.63 & 0.66 \\ 
  12 & 1206.54 & 0.16 & 12.31 & 0.55 & 1334.56 & 0.15 & 10.26 & 0.51 & 1335.70 & 0.20 & 18.54 & 0.64 \\ 
  13 & 1206.51 & 0.20 & 10.98 & 0.53 & 1334.56 & 0.09 & 9.19 & 0.49 & 1335.70 & 0.22 & 17.30 & 0.62 \\ 
  14 & 1206.48 & 0.25 & 19.60 & 0.65 & 1334.55 & 0.12 & 12.50 & 0.55 & 1335.69 & 0.22 & 22.31 & 0.69 \\ 
  14a & 1206.50 & 0.17 & 12.22 & 1.34 & 1334.57 & 0.12 & 10.20 & 1.16 & 1335.70 & 0.23 & 17.75 & 1.40 \\ 
  14b & 1206.46 & 0.34 & 24.29 & 1.57 & 1334.55 & 0.18 & 13.70 & 1.27 & 1335.68 & 0.13 & 24.51 & 1.54 \\ 
  14c & 1206.48 & 0.33 & 21.79 & 1.31 & 1334.55 & 0.12 & 13.39 & 1.08 & 1335.68 & 0.21 & 24.22 & 1.33 \\ 
  15 & 1206.53 & 0.13 & 12.01 & 0.55 & 1334.58 & 0.11 & 9.29 & 0.51 & 1335.71 & 0.16 & 18.59 & 0.64 \\ 
  16 & 1206.53 & 0.12 & 12.22 & 0.55 & 1334.57 & 0.13 & 9.89 & 0.50 & 1335.69 & 0.21 & 17.83 & 0.63 \\ 
  17 & 1206.50 & 0.18 & 12.49 & 0.56 & 1334.56 & 0.12 & 9.37 & 0.49 & 1335.70 & 0.24 & 18.21 & 0.63 \\ 
  18 & 1206.52 & 0.16 & 12.47 & 0.56 & 1334.58 & 0.12 & 9.70 & 0.50 & 1335.71 & 0.21 & 18.44 & 0.64 \\ 
  19 & 1206.53 & 0.18 & 11.47 & 0.54 & 1334.57 & 0.17 & 8.83 & 0.50 & 1335.71 & 0.20 & 18.60 & 0.64 \\ 
  20 & 1206.54 & 0.17 & 10.56 & 0.53 & 1334.56 & 0.12 & 9.07 & 0.49 & 1335.70 & 0.22 & 17.65 & 0.63 \\ 
\hline
  \end{tabular}
}
\end{table}

\clearpage

\begin{table}
\centering
\caption{\label{lineflu3} 
Line measurements as in Table \ref{lineflu1} for lines of C III multiplet and N V doublet.} 
\resizebox{0.99\textwidth}{!}{
\begin{tabular}{l|rrrr|rrrr|rrrr}
  \hline
  \hline
Exposure & Wavelength & FWHM & Flux & Error & Wavelength & FWHM & Flux & Error & Wavelength & FWHM & Flux & Error \\ 
         & \AA\  & \AA\ & \multicolumn{2}{c|}{$10^{-15}$ erg cm$^{-2}$ s$^{-1}$} &  \AA\  & \AA\ & \multicolumn{2}{c|}{$10^{-15}$ erg cm$^{-2}$ s$^{-1}$} &  \AA\  & \AA\ & \multicolumn{2}{c}{$10^{-15}$ erg cm$^{-2}$ s$^{-1}$} \\
\hline
 & \multicolumn{4}{c|}{C III \@ 1175 \AA\ } & \multicolumn{4}{c|}{N V \@ 1238.8 \AA\ } & \multicolumn{4}{c}{N V \@ 1242.7 \AA\ } \\
  1 & 1175.71 & 0.85 & 10.21 & 0.96 & 1238.81 & 0.19 & 3.40 & 0.51 & 1242.73 & 0.32 & 2.05 & 0.47 \\ 
  2 & 1175.72 & 0.74 & 9.67 & 0.94 & 1238.81 & 0.12 & 3.54 & 0.50 & 1242.75 & 0.08 & 2.14 & 0.46 \\ 
  3 & 1175.71 & 0.96 & 9.17 & 0.93 & 1238.81 & 0.14 & 3.53 & 0.50 & 1242.71 & 0.16 & 2.27 & 0.47 \\ 
  4 & 1175.72 & 0.21 & 10.95 & 0.96 & 1238.84 & 0.18 & 3.19 & 0.50 & 1242.72 & 0.11 & 2.24 & 0.47 \\ 
  5 & 1175.73 & 0.96 & 36.63 & 1.02 & 1238.83 & 0.16 & 8.03 & 0.45 & 1242.76 & 0.15 & 4.54 & 0.39 \\ 
  5a & 1175.73 & 1.56 & 64.92 & 2.71 & 1238.85 & 0.36 & 11.20 & 1.18 & 1242.76 & 0.16 & 6.63 & 1.07 \\ 
  5b & 1175.70 & 0.22 & 25.84 & 2.14 & 1238.82 & 0.09 & 7.21 & 1.09 & 1242.79 & 0.09 & 3.96 & 1.00 \\ 
  5c & 1175.75 & 1.03 & 22.07 & 1.79 & 1238.82 & 0.12 & 6.06 & 0.91 & 1242.72 & 0.16 & 3.28 & 0.83 \\ 
  6 & 1175.73 & 0.46 & 14.92 & 0.76 & 1238.80 & 0.25 & 4.95 & 0.39 & 1242.78 & 0.18 & 2.63 & 0.34 \\ 
  7 & 1175.73 & 0.82 & 12.43 & 0.73 & 1238.80 & 0.24 & 4.10 & 0.37 & 1242.76 & 0.21 & 2.26 & 0.34 \\ 
  8 & 1175.73 & 0.18 & 10.86 & 0.71 & 1238.83 & 0.15 & 3.99 & 0.38 & 1242.73 & 0.22 & 2.35 & 0.34 \\ 
  9 & 1175.67 & 0.20 & 9.39 & 0.69 & 1238.82 & 0.22 & 3.44 & 0.36 & 1242.67 & 0.19 & 2.32 & 0.34 \\ 
  10 & 1175.74 & 1.41 & 10.54 & 0.71 & 1238.81 & 0.23 & 3.75 & 0.37 & 1242.72 & 0.13 & 2.00 & 0.33 \\ 
  11 & 1175.67 & 0.33 & 12.84 & 0.74 & 1238.78 & 0.24 & 4.27 & 0.38 & 1242.80 & 0.09 & 2.41 & 0.34 \\ 
  12 & 1175.70 & 1.48 & 10.33 & 0.70 & 1238.82 & 0.22 & 3.75 & 0.38 & 1242.69 & 0.98 & 1.90 & 0.33 \\ 
  13 & 1175.68 & 1.11 & 9.70 & 0.69 & 1238.81 & 0.19 & 3.16 & 0.36 & 1242.71 & 0.18 & 1.81 & 0.33 \\ 
  14 & 1175.60 & 1.39 & 29.34 & 0.95 & 1238.77 & 0.21 & 5.70 & 0.41 & 1242.76 & 0.20 & 3.42 & 0.36 \\ 
  14a & 1175.68 & 1.45 & 11.74 & 1.89 & 1238.77 & 0.09 & 3.09 & 0.98 & 1242.78 & 0.67 & 2.13 & 0.95 \\ 
  14b & 1175.59 & 1.27 & 43.86 & 2.45 & 1238.76 & 0.17 & 6.59 & 1.08 & 1242.74 & 0.44 & 4.50 & 1.01 \\ 
  14c & 1175.58 & 0.27 & 31.85 & 1.92 & 1238.77 & 0.33 & 7.09 & 0.91 & 1242.76 & 0.25 & 3.60 & 0.82 \\ 
  15 & 1175.72 & 0.08 & 13.33 & 0.75 & 1238.80 & 0.21 & 3.99 & 0.38 & 1242.75 & 0.28 & 2.49 & 0.34 \\ 
  16 & 1175.69 & 0.22 & 10.81 & 0.71 & 1238.83 & 0.14 & 3.91 & 0.38 & 1242.75 & 0.23 & 2.25 & 0.34 \\ 
  17 & 1175.70 & 0.12 & 10.90 & 0.71 & 1238.79 & 0.15 & 3.99 & 0.37 & 1242.76 & 0.16 & 2.24 & 0.34 \\ 
  18 & 1175.73 & 0.13 & 11.03 & 0.72 & 1238.80 & 0.08 & 3.85 & 0.37 & 1242.74 & 0.12 & 2.57 & 0.34 \\ 
  19 & 1175.70 & 1.37 & 10.05 & 0.70 & 1238.80 & 0.19 & 3.74 & 0.37 & 1242.75 & 0.18 & 2.09 & 0.33 \\ 
  20 & 1175.69 & 0.11 & 9.53 & 0.69 & 1238.81 & 0.16 & 3.48 & 0.37 & 1242.75 & 0.14 & 2.02 & 0.33 \\ 
  \hline
  \end{tabular}
}
\end{table}



{\it Facilities:} \facility{HST (COS)}.


\end{document}